# Fine-Tuning Topics through Weighting Aspect Keywords

Ali Nazari[1]*, Michael Weiss[2]

[1] Sprott School of Business, Carleton University, Ottawa, ON, Canada
[2] Technology Innovation Management (TIM) program, Carleton University, Ottawa, ON, Canada

*Corresponding author(s): ali.nazari@carleton.ca
Contributing authors: michael.weiss@carleton.ca

## Abstract

Organizations face growing challenges in deriving meaningful insights from vast amounts of specialized text data. Conventional topic modeling techniques are typically static and unsupervised, making them ill-suited for fast-evolving fields like quantum cryptography. These models lack contextual awareness and cannot easily incorporate emerging expert knowledge or subtle shifts in subdomains. Moreover, they often overlook rare but meaningful terms, limiting their ability to surface early signals or align with expert-driven insights essential for strategic understanding. To tackle these gaps, we employ design science research methodology to create a framework that enhances topic modeling by weighting aspects based on expert-informed input. It combines expert-curated keywords with topic distributions iteratively to improve topic relevance and document alignment accuracy in specialized research areas. The framework comprises four phases, including (1) initial topic modeling, (2) expert aspect definition, (3) supervised document alignment using cosine similarity, and (4) iterative refinement until convergence. Applied to quantum communication research, this method improved the visibility of critical but low-frequency terms. It also enhanced topic coherence and aligned topics with the cryptographic priorities identified by experts. Compared to the baseline model, this framework increased intra-cluster similarity. It reclassified a substantial portion of documents into more thematically accurate clusters. Evaluating QCrypt 2023 and 2024 conference papers showed that the model adapts well to changing discussions, marking a shift from theoretical foundations to implementation challenges. This study illustrates that expert-guided, aspect-weighted topic modeling boosts interpretability and adaptability. It also enhances strategic relevance, providing a scalable solution for tracking knowledge in fast-paced research domains.

## 1. Introduction

In fast-evolving scientific and technological fields, organizations increasingly struggle to extract timely, context-aware insights from large volumes of domain-specific literature. Although topic modeling is widely used to uncover latent patterns in unstructured text, conventional techniques, such as Latent Dirichlet Allocation (LDA) [1] and clustering-based approaches like BERTopic [2], are often unsupervised, static, and disconnected from expert perspectives. These models prioritize dominant statistical patterns and frequently overlook rare or low-frequency terms that may signal emerging trends or carry strategic importance.

This limitation is especially problematic in domains where subtle, infrequent concepts may reflect critical innovation trajectories or weak signals of change. Simply increasing the number of topics to improve coverage often leads to fragmented, redundant outputs that hinder interpretability. Moreover, these models typically lack the capacity to reflect evolving expert knowledge, resulting in reduced alignment with real-world priorities and missed opportunities for actionable insight. As a result, organizations using standard topic models often encounter challenges related to limited interpretability, poor adaptability, and an inability to capture the nuanced evolution of domain-specific discourse [3], [4].

To address these shortcomings, several enhancements to topic modeling have been proposed. These include weighted keyword extraction [5], aspect-oriented clustering [6], and semi-supervised interpretation frameworks [7], [8]. While such techniques improve topic coherence and interpretability, they often rely on static training data, predefined taxonomies, or ex-post validations. Moreover, few approaches provide dynamic mechanisms for integrating expert input during model formation, a major limitation when decision-making depends on detecting emerging trends and weak signals [9], [10].

To address this gap, we proposed an Aspect-Weighted Topic Modeling method. This approach iteratively integrates expert-informed aspect keywords into topic modeling workflows. By using curated and weighted keyword sets drawn from domain expertise, we score and refine topic-document associations using similarity metric. Our method supports iterative reassignment of documents based on aspect relevance, allowing rare but strategically significant terms to influence topic boundaries and improving coherence over time. The research addresses the following question: How can external expert knowledge be systematically integrated into topic modeling to improve contextual alignment and adaptability in specialized domains? To answer this, we develop and evaluate a multi-step method that includes literature mapping, aspect construction, iterative topic refinement, and expert-guided evaluation.

This study makes three key contributions. The proposed method is designed for allows for adaptability, enabling continuous topic refinement through the incorporation of new expert-informed inputs and capturing domain shifts in fast-moving, innovation-driven fields. This creates an automated and contextually tuned updating mechanism. Besides, the method's applicability is demonstrated in a real-world technological domain, where it helps detect weak signals, identify emerging research trends, and support foresight in fields such as quantum communication.

To validate the framework, we apply it to a curated corpus of academic publications and patents in the quantum communication cryptography domain. This field is ideal for testing our method due to its rapid evolution, technical complexity, and use of emerging terminology. Evaluation is performed through quantitative metrics (e.g., topic coherence, document assignment).

In the following sections, we review related work, present the design and implementation of our method using DSRM, and evaluate it through the quantum communication case study. This evaluation demonstrates how the integration of expert-defined aspects supports weak signal detection and improved document clustering.

## 2. Related Work

We conducted a structured literature review following the systematic mapping methodology proposed by Petersen et al. (2008 and 2015) to support the development of our framework and identify methodological gaps in topic modeling. Our process included keyword-based article retrieval, TF-IDF-based vectorization, cosine similarity analysis, and hierarchical clustering, ultimately yielding five thematic clusters. These clusters, labeled S1 through S5, are derived by analyzing how various approaches contribute to topic modeling and document clustering across four guiding sub-questions (Q1.1 to Q2.2). As shown in Figure 1, this review structure enabled us to link each research question to one thematic strands, which in turn informed the design of the framework's four core components: topic modeling, aspect-weighted keywords, supervised alignment, iterative refinement, and a future integration of reinforcement learning. The five thematic strands are summarized as follows:

S1: Topic Modeling - Studies that explore the unsupervised extraction of latent themes.
S2: Topic Examination - Research is for refining topic models through keyword weighting or structural enhancements.
S3: Multi-faceted Topic Insights - methods that enable expert-guided interpretation of overlapping or layered topics.
S4: Supervised/Semi-Supervised Clustering - These approaches revolve around using classified data or anchor concepts to influence topic emergence.
S5: Deep Learning and Reinforcement Learning - Advanced models and frameworks that incorporating adaptive mechanisms and learning feedback.

These stands together emphasize methodological shortcomings in the current literature: static topic structures, limited contextual keyword weighting, minimal expert feedback, and infrequent use of dynamic, reward-driven topic adaptation. Our framework tackles these gaps by aligning its components.

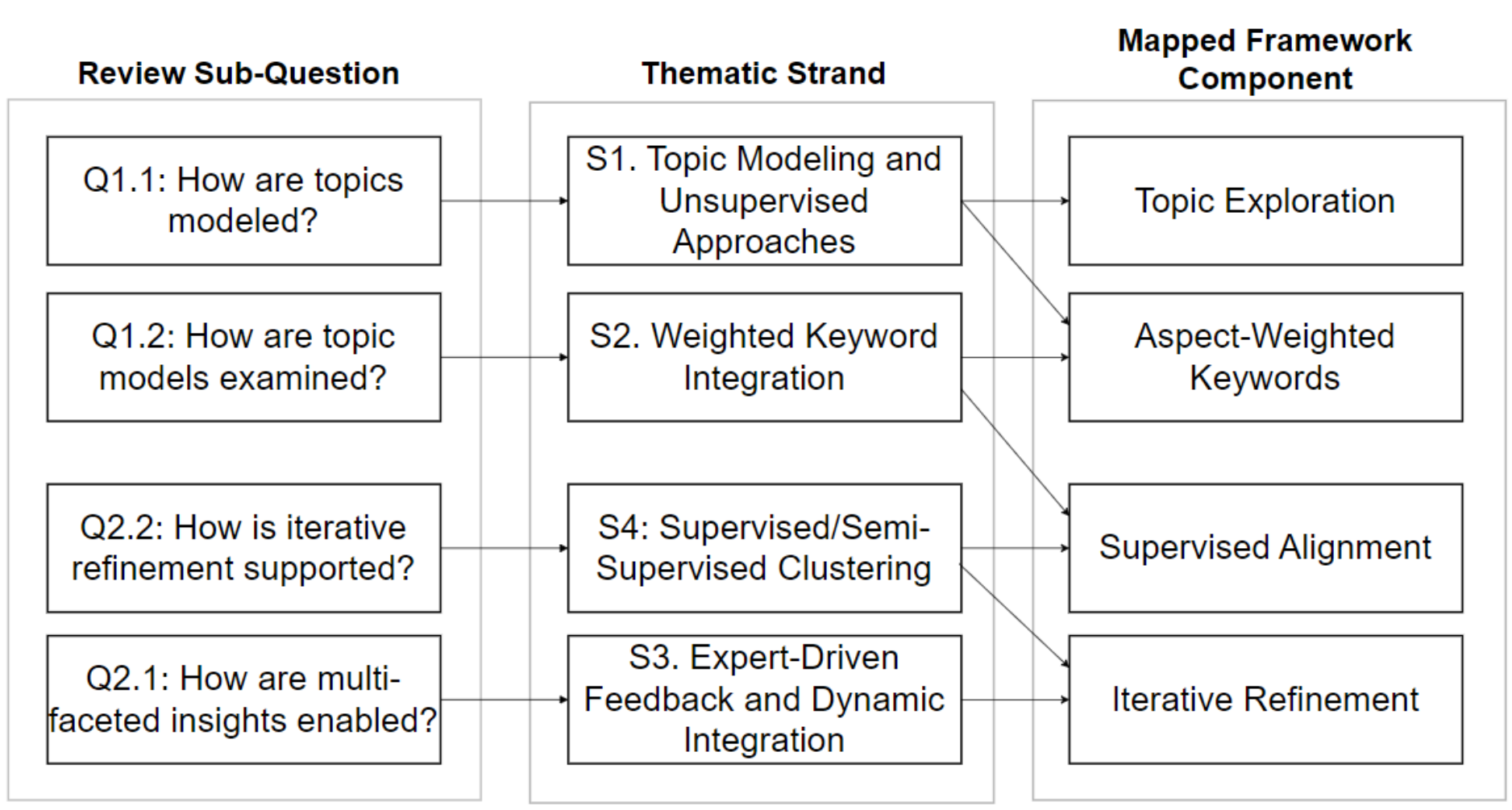

**Fig 1:** Mapping of Literature Review Sub-Questions (Q1.1–Q2.2) to Thematic Strands (S1–S5) and Mapped Framework Components

This figure illustrates the structured flow from four review sub-questions (Q1.1–Q2.2) through five thematic strands (S1–S5), culminating in the design of four core components of the proposed framework: aspect-weighted keywords, supervised alignment, iterative refinement, and future reinforcement learning integration. While S5 emerged from the literature, it is not implemented in this study and is excluded from the figure. The following subsections examine each thematic strand (S1–S5) in detail (see some papers in Appendix 1), beginning with unsupervised topic modeling approaches.

### *Topic Modeling and Unsupervised Approaches (S1)*

Latent Dirichlet Allocation (LDA) [1] is the foundational technique for topic modeling. LDA extensions such as dynamic topic models [13] and coherence-improving methods [8] enhance topic stability and interpretability. However, most methods remain static and frequency-based. They lack contextual adaptability and domain-specific relevance. In the domain of quantum communication, [14] showed the utility of LDA to uncover topic trends, but also highlighted its limitations in early signal detection. [7] stated a fuzzy clustering approach to improve online document classification, while [5] showed how short-text clustering benefits from deep self-monitoring models. Still, these approaches struggle to adapt to fast-evolving corpora where new terms emerge quickly.

### *Weighted Keyword Integration (S2)*

Some studies have utilized keyword weighting techniques such as TF-IDF and word embeddings [15] to improve precision in topic identification. [5] showed how weighting can improve short-text clustering performance, and aspect-driven keywords to improve classification accuracy.

In the realm of cryptography and other niche areas, TF-IDF-based techniques have been extensively employed to enhance the depiction of infrequent yet semantically significant terms, a fundamental concept presented by [16] in the framework of term-weighting for data retrieval. Nonetheless, most existing methods rely on pre-defined heuristics rather than dynamic weighting, failing to reflect evolving expert priorities and [17], [18] both explored methods to refine topic structures over time using fuzzy or supervised integration, but lacked expert-driven control mechanisms.

### *Expert-Driven Feedback and Dynamic Integration (S3)*

Human-in-the-loop systems are gaining attention for improving topic relevance. [8] proposed iterative feedback mechanisms using expert annotations. [19] and [20] demonstrated how seeded terms can act as anchors for dynamic model correction. These approaches show promise, especially in fields like quantum communication where expert oversight is critical.

Most methods use expert input post hoc rather than embedding it during training. [21] introduced dual-supervised clustering. They used feature re-weighting to improve interpretability. [22] applied NLP-guided semi-supervised clustering. Their work aimed to reveal hidden structures in research proposals but lacked dynamic feedback loops. The literature shows a clear need for ongoing expert involvement in adapting models.

### *Supervised and Semi-Supervised Clustering (S4)*

Seeded and guided topic models offer a compromise between manual annotation and unsupervised modeling. SeededLDA [19] and similar guided approaches leverage expert-labeled terms to direct topic formation. [6] demonstrated improved coherence using supervised and semi-supervised combinations. [23] proposed constraint generation to refine document clustering. [24] explored document clustering in collaborative workspaces, adapting to report structures. However, these models often require extensive labeled data and struggle in data-sparse, rapidly evolving domains where supervision is limited.

### *Reinforcement Learning for Topic Adaptation (S5)*

Recent work highlights the potential of reinforcement learning (RL) for adaptive topic modeling. [25] proposed a feedback-based system where topic relevance is adjusted using downstream classification accuracy. [26] applied RL to refine topic boundaries over time, but focused primarily on guiding topic models to detect weak signals in dynamic corpora. [27] introduced a multi-level supervised neural topic model, while [28] used metric learning to detect diverse topic viewpoints. [8] combined graph convolutional networks with clustering to handle dense, overlapping topic spaces. Despite these advancements, RL frameworks rarely integrate expert-informed signals into their reward functions, a limitation that our proposed system addresses in future work. The technical complexity and limited data availability in emerging domains like quantum cryptography further constrained RL's immediate feasibility. For this reason, S5 is acknowledged as an important future direction but is excluded from the current framework and Figure 1, which focuses on the implemented components.

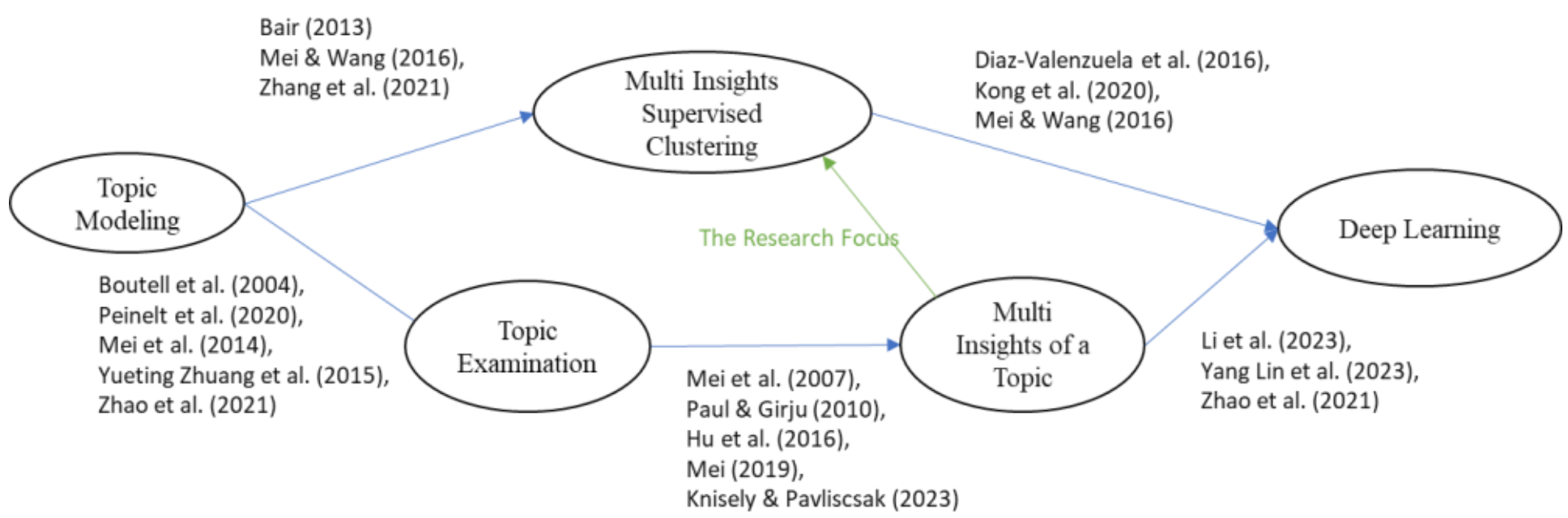

**Fig 2:** Bridging Topic Modeling to Deep Learning for Document Analysis

Following the thematic classification shown in Figure 2, we examined representative papers from each of the five clusters (S1–S5) to understand how current research addresses topic modeling, clustering, and document understanding. This figure complements Figure 1 by showing how traditional modeling, expert feedback, and reinforcement learning are connected across the literature.

### *Summary and Research Gaps*

The reviewed literature presents a rich set of innovations across unsupervised, semi-supervised, and adaptive modeling. However, several key limitations persist, as summarized in Table 1:

**Table 1:** Summary of research gaps relevant to this study

| Gap Type | Description | Key References |
|---|---|---|
| Methodological | Topic models lack contextual integration and rely on static configurations, limiting adaptability and relevance. | [3], [4], [29] |
| Practical | Expert input is not dynamically or systematically integrated, reducing its strategic impact in domains like cryptography. | [9], [10] |
| Theoretical | Few models support continuous learning or reward-driven mechanisms to optimize topic relevance over time. | [25], [26] |

This structured approach enhances weak signal detection and enables dynamic topic alignment, while laying the foundation for future integration of reward-based mechanisms such as reinforcement learning.

## 3. Research Design and Methodology

This research utilizes the Design Science Research Methodology (DSRM) [30], [31] to create and evaluate our proposed framework as an artifact. To ensure design traceability, Figure 1 is revisited to demonstrate how each phase of the proposed framework directly corresponds to thematic strands (S1–S5) and addresses specific methodological gaps identified through the literature review. This mapping shows that the framework was not built arbitrarily but emerged systematically from a structured synthesis of prior research, aligning sub-questions, empirical findings, and system design requirements.

The proposed framework incorporates expert-informed aspect keywords in a systematic manner to enhance topic coherence and document clustering in domain-specific corpora. We follow the six steps of the DSRM approach for creating purposeful artifact that include:

1. Problem Identification: This research addresses limitations in topic modeling techniques which struggle to detect low-frequency but important signals in specialized areas. Current methods, like Latent Dirichlet Allocation (LDA) extensions or models such as BERTopic, often create coherent topics. However, they do not align with the evolving concerns and priorities recognized by domain experts. This gap is critical in fields like quantum cryptography, where the timely detection of emerging themes is crucial.
2. Objectives: The goal is to build an artifact that integrates expert knowledge to improve topic relevance and interpretability, enabling dynamic detection of emerging or subtle aspects across document collections.
3. Artifact Design: A framework is designed to combine topic modeling, expert-defined aspect keywords, and supervised clustering, around four components:
   (1) baseline topic modeling,
   (2) expert-informed aspect integration,
   (3) document-aspect alignment via similarity, and
   (4) iterative refinement for coherence.
4. Demonstration: The framework is tested on a curated corpus in quantum communication, using domain expertise (e.g., conference topics) to guide and evaluate refinement.
5. Evaluation: Performance is assessed both quantitatively (topic coherence) and qualitatively (thematic alignment), supported by heatmaps, clustering visuals, and keyword relevance.
6. Communication: The framework contributes to literature on semi-supervised topic modeling and offers practitioners a scalable, interpretable tool for detecting technological change. Its modular design supports transfer to other domains.

The next chapter details how each component of the framework is implemented to address the design requirements and solve the core problem.

## 4. Design and Development of the Framework

Our framework is built around four key components, each representing a distinct phase in a structured, step-by-step process. These phases are designed to meet specific design requirements, which are outlined in the following subsection.

### 4.1. Design Requirements

Drawing from the problem context and our research objectives, we identified five core design requirements to guide how the method, developed within this broader framework, should function:

DR1: Support weak signal detection: The method needs to capture rare but meaningful terms that typical frequency-based topic models tend to miss (see Section 4.5 on Aspect Keyword Weighting).

DR2: Integrate expert knowledge: It should bring in external insights by using domain-specific aspect keywords, sourced from conference literature, as a practical stand-in for expert input (see Section 4.5.2 on Aspect Definition & Scoring).

DR3: Strengthen document-topic alignment. Topics should align with expert-identified themes, not follow word frequency patterns (see Section 4.6 on Supervised Clustering and Refinement).

DR4: Ensure interpretability: The topic structures must be clear and meaningful. They should help experts review and detect technological shifts (see Section 4.4 on Initial LDA and Realignment).

DR5: Enable iterative refinement. This method should allow for ongoing tuning and adjustments. This will help improve the clarity and stability of the topics over time (see Section 4.7 on Iterative Refinement).

## 4.2. Core Components of the proposed method

The proposed framework is a modular and iterative system. It fine-tunes topic models with guidance from experts. Its structure is based on five key design requirements (DR1–DR5). These focus on detecting weak signals, incorporating expert knowledge, improving document alignment with topics, ensuring interpretability, and supporting ongoing refinement (Figure 3). In the next sections, we explain the design thinking for each core component, explaining how this method is tailored to meet the unique challenges of analyzing texts in fast-evolving fields.

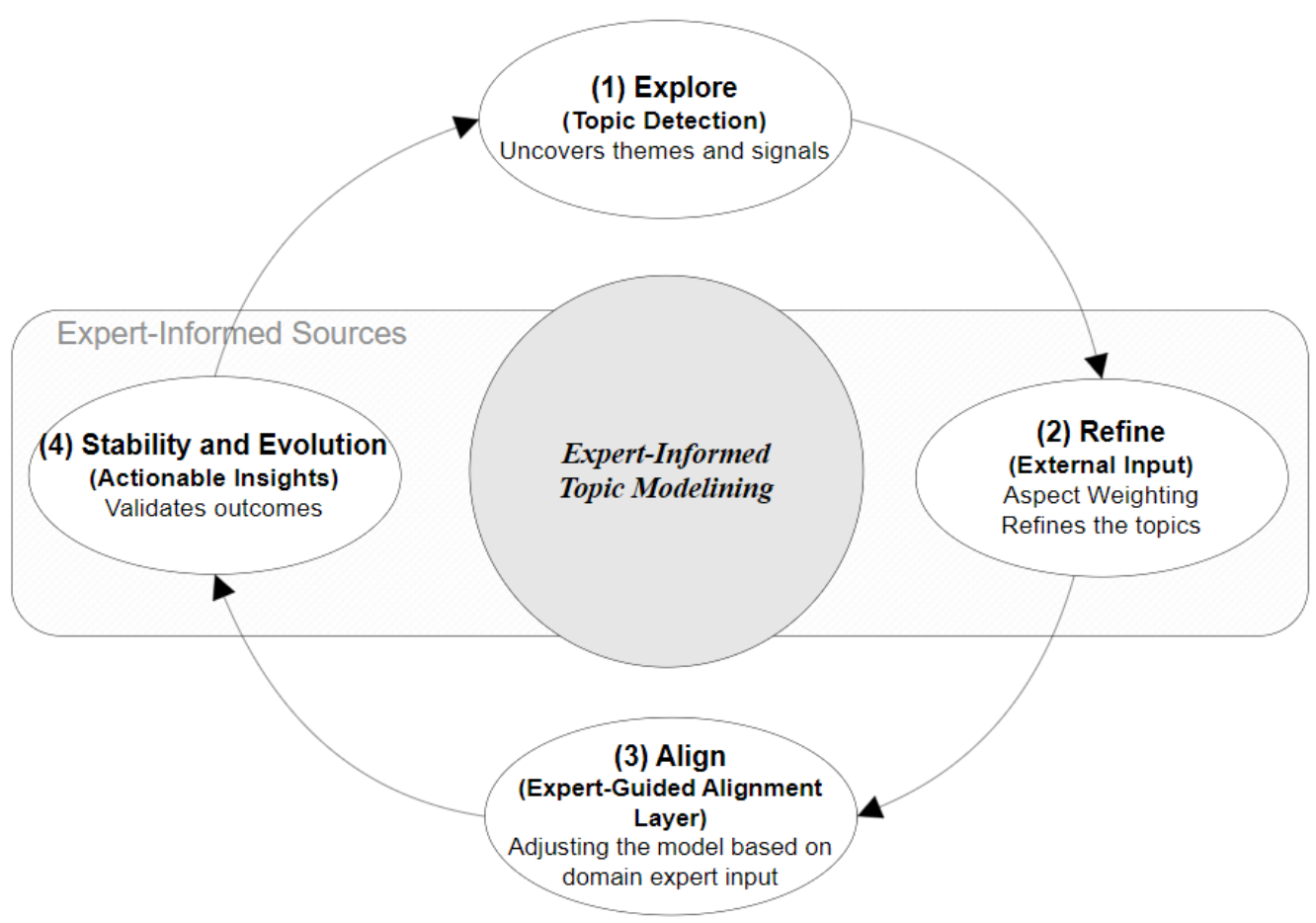

**Fig 3:** Expert-Informed Iterative Framework for Topic Refinement and Alignment

Component 1: Initial Topic Modeling - This component use a unsupervised model offers a flexible baseline. It helps capture broad thematic structures. It supports two requirements, DR1 (weak signals) and DR4 (interpretability). It uses the model as an initial categorization.

Component 2: Aspect Definition Layer - This layer uses curated aspect vectors to embed expert knowledge. These vectors show strategic subdomains through weighted terms. This addresses DR2 (expert input) and enhances DR1, allowing meaningful low-frequency terms to shape topic interpretation.

Component 3: Document Realignment via Semantic Scoring - this component reassigns documents by how closely they match expert-defined aspects. This is done without retraining the model. This supervised overlay matches documents to domain priorities. It meets DR3 (alignment) and helps DR4 by creating clearer clusters.

Component 4: Iterative Refinement - A feedback loop adjusts topic-aspect alignments. This process happens over iterations, enhancing model coherence and adaptability. This supports DR5 (refinement) by enabling the system to evolve with emerging signals and domain shifts.

The method creates a layered structure. It combines machine-driven discovery with human guidance. This balance improves topic usefulness and strategic insight.

## 4.3. Implementation and Case Study: Quantum Communication Domain

We applied the Aspect-Weighted Topic Modeling framework in quantum communication. This section shows how we implemented the framework components, as seen in Figure 4, using a step-by-step workflow in Figure 5. Each implementation phase matches a framework component. This ensures we follow the design logic from initial topic extraction to iterative refinement.

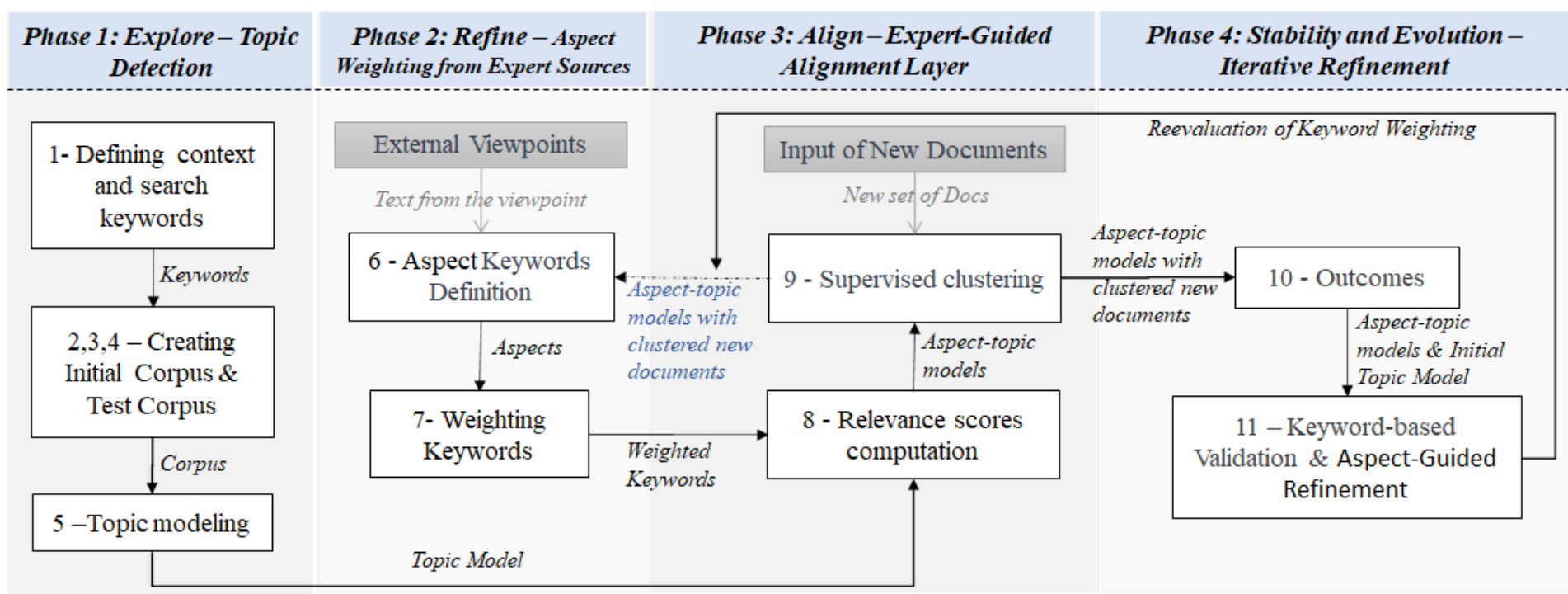

** The numbers correspond to line numbers in the pseudocode.*

**Fig 4:** Overview of the Aspect-Weighted Topic Modeling Method workflow, showing the multi-phase process from data preparation to topic refinement.

### *Case Domain and Corpus Construction*

The case domain is selected based on its strategic importance and the presence of high-velocity scientific output. Quantum communication, including quantum key distribution (QKD), post-quantum cryptography, and entanglement-based networking, represents an area where traditional topic modeling often fails to capture nuanced developments due to rapidly shifting terminologies and low-frequency innovations. A corpus of 1,048 documents was curated from peer-reviewed journals and conference proceedings indexed in Web of Science and Scopus, covering the years 2017–2024, and publications of a conference 2023 and 2024 as test of the proposed framework. This selection prioritized high-impact publications and conference outputs known for surfacing frontier research (QCrypt 2023[3], QCrypt 2024[4], Quantum.Tech[5]).

### *Framework Application Overview*

The method was applied in four phases within the quantum communication domain:

Phase 1: Topic Detection - An LDA model uncovered 39 latent topics. While they noted themes like "quantum networks," many struggled to identify new cryptographic concerns clearly.

[3] https://qcrypt.net/2023/accepted-papers/
[4] https://qcrypt.net/2024/accepted-papers/
[5] https://www.alphaevents.com/events-quantumtechus

Phase 2: Aspect Weighting from Expert Sources - Aspect keyword sets from Quantum. Tech proceedings used TF-IDF to encode strategic themes. This uncovered rare but important terms that the baseline model missed.

Phase 3: Expert-Guided Alignment - We compare documents to aspect vectors by calculating cosine similarity. After that, we assign each document to the most closely aligned aspect. This improved specificity divides broad topics into clearer subfields like “device-independent QKD” and “multi-photon interference.”

Phase 4: Iterative Refinement - document-aspect assignments evaluate until they converge. This boosts similarity within clusters and improves expert-rated interpretability. Final clusters are aligned with emerging themes like post-quantum migration and hardware integration.

### *Observations and Practical Implications*

The method is effective in:

(1) Improving interpretability: Expert-guided alignment showed how experts view subfields.

(2) Amplifying weak signals: We highlighted low-frequency but important terms using aspect weighting.

(3) Achieving semantic convergence: After repeated refinement, we created stable clusters that are coherent and thematically relevant.

These results show the value of integrating expert insights in document-topic modeling. A pseudocode overview (Appendix 6) details the full pipeline aligned with four phases shown in Figure 5, from topic modeling to alignment and guided clustering. Sections 4.4–4.7 explain each phase with examples from the quantum communication case study.

## 4.4. Phase 1: Explore – Topic Detection

We built the corpus by retrieving peer-reviewed papers, patents, conference proceedings, and whitepapers from Web of Science and Scopus using a focused query on quantum communication. After preprocessing (tokenization, lemmatization, stop-word removal, and normalization), we applied LDA for topic modeling. Based on C-V coherence scores (Figure 5), six primary topics were identified and refined into 39 subtopics (T1–T39) using hierarchical clustering (Figure 6).

### *Data Collection*

We focus on the quantum communication domain to demonstrate the method’s effectiveness, specifically, cryptography, quantum key distribution (QKD), and quantum networks. Using these focal areas, we compiled a corpus of academic papers from online libraries that reflect current research and emerging trends.

### *Defining context and search keywords*

Quantum communication is critical for sectors like technology, finance, and aerospace. The 2023–2024 Quantum.Tech conferences helped guide our selection of relevant aspects and keywords, including quantum, network, communication, development, application, experiment, and algorithm. These keywords were derived from seminal papers [32], [33], [34] and refined using preprocessing techniques such as stop-word removal, stemming, lemmatization, and word cloud analysis.

The final search string used was:

"quantum AND (communication OR network) AND (development OR application OR experiment OR implement OR algorithm OR use) AND (feasibility OR future OR forecast OR trend OR progress)"

Documents were retrieved from Scopus and Web of Science, focusing on peer-reviewed articles, reviews, and conference papers. To expand the dataset, we used Litmaps, a citation network tool, to extract and add related documents.

### *Corpus*

The process of creating a corpus has two sub-steps, searching and screening documents in online libraries.

#### *Search Step*

A total of 3,527 documents were retrieved (1,600 from Web of Science, 1,927 from Scopus). After applying relevance filters, 2,406 documents were retained, drawn from leading journals such as IEEE Access, Optics Express, and Proceedings of SPIE.

#### *Screening Step*

To identify the most relevant documents, we used a neural network classifier trained on a labeled sample of 120 documents (60 relevant, 60 irrelevant), using uncertainty sampling [35]. Features were based on word frequencies. This process selected 1,048 high-relevance documents for topic modeling. Documents were then preprocessed through: tokenization, stemming and lemmatization, stop-word and special character removal, and lowercasing and URL filtering. Metadata such as document ID, title, abstract, DOI, year are preserved for modeling.

#### *Summary*

Search results: 3,527 documents → filtered to 2,406 relevant papers

Screened corpus: 1,048 relevant documents identified using neural network classification

With the curated and preprocessed corpus in place, we proceeded to apply topic modeling to uncover latent themes and structure within the quantum communication research landscape.

### *Topic Model Analysis*

We applied Latent Dirichlet Allocation (LDA) to the 1,048 preprocessed documents to uncover latent themes. The model revealed six primary topics (named PTs), with the optimal number determined using C-V coherence scores [1], which assess the semantic similarity among top words in each topic (Figure 6).

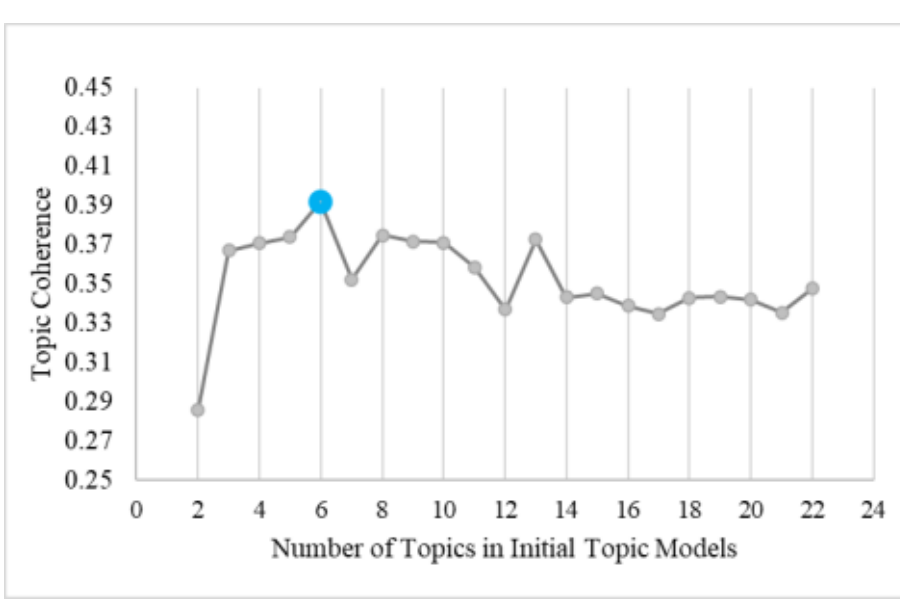

**Fig 5:** Coherence Scores for Various Numbers of Primary Topics

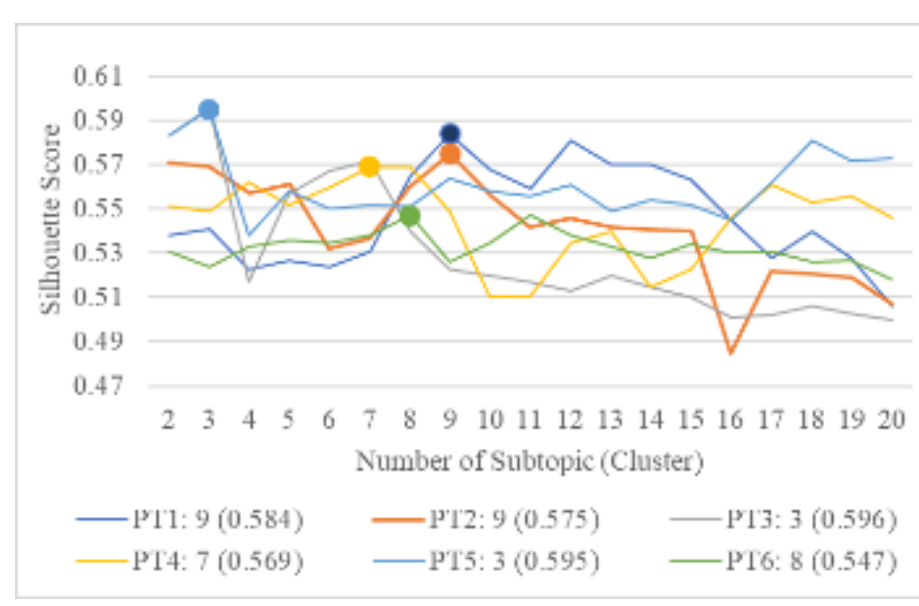

**Fig 6:** Silhouette Scores for Various Numbers of SubTopics

To improve coherence, we applied hierarchical clustering within each primary topic, creating finer-grained subtopics. The number of subtopics is optimized using silhouette scores, which evaluate how well a document fits its assigned cluster versus others:

$$s = \frac{b-a}{max(a,b)} \text{ (1)},$$

where $a$ is the average distance from a data point to all others in the same cluster (cohesion). b is the average distance to all points in the nearest neighboring cluster (separation). We calculated silhouette scores for different numbers of clusters. This found the optimal number of subtopics for each initial topic. The greatest number of subtopics for each primary topic (Figure 7) is 9 for PT1, 9 for PT2, 3 for PT3, 7 for PT4, 3 for PT5, and 8 for PT6. This gives a total of 39 subtopics. From now on, we will refer to these subtopics as topics from T1 to T39 for ease of use. Figure 8 presents the graph depicting these topics, along with the top three words for each.

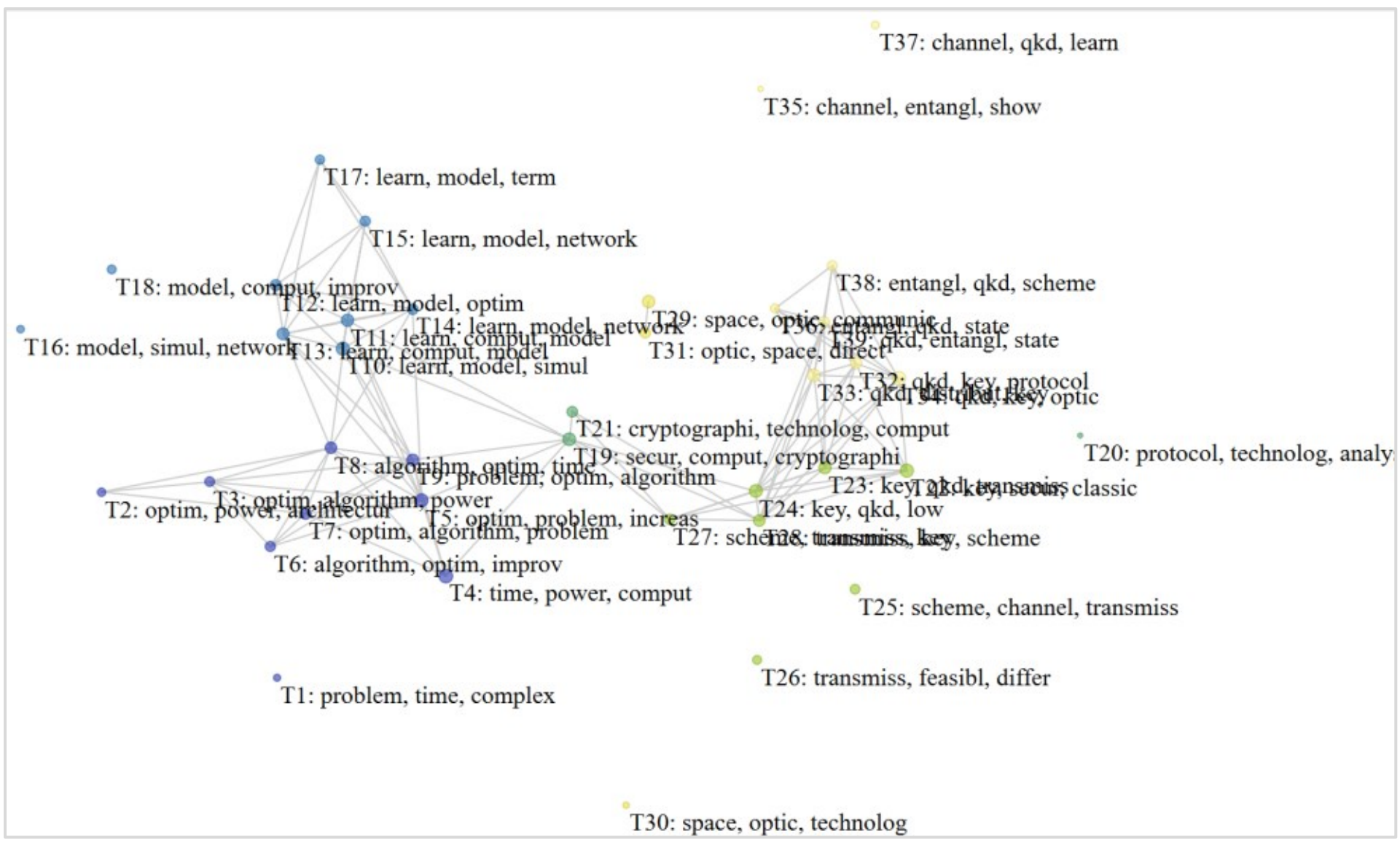

**Fig 7:** the network of the topics

Figure 8 illustrates the 39 topics (T1–T39) derived from hierarchical clustering, each labeled with its top three keywords to summarize its main theme. Lines between topics indicate their similarity, while the size of filled circles represent topic weights, with the six topic colors, reflecting their importance in the corpus. In later analysis, we examine how aspect emphasis influences these weights to reveal deeper structural relationships among the topics.

*Summary of topic modeling step*

Preprocessing Output: A cleaned and standardized corpus of 1,048 documents ready for topic modeling.

Topic Modeling Output: LDA found 6 topics. C-V scores set the optimal number. We refined the 39 subtopics from the primary topics using hierarchical clustering and labeled them T1 to T39.

## 4.5. **Phase 2:** Refine – Aspect Weighting from Expert Sources

We identified nine strategic domain aspects, Cryptography, Networks, Tech Research, QKD, Entanglement, Teleportation, Channels, Repeaters, and Applications by analyzing curated sources such as Quantum.Tech 2023–2024 (Section 4.5.1). For evaluation, we focus on the Cryptography aspect and then two more Cryptography advancements to apply to topic models in two iterations. We collected weighted keyword lists from the conference proceedings and a proxy "Quantum Tech Report".

### *Aspect Identification*

Aspect corpora are constructed using expert sources (e.g., conference proceedings like Quantum.Tech papers) aligned with current quantum communication discourse. These included documents published after the main corpus date, ensuring relevance to evolving trends. Each aspect (e.g., Cryptography, Networks) is represented by a curated document set, forming an "aspect corpus." From these, we extracted the top 50 TF-IDF-weighted keywords per aspect (Section 4.5.2), creating aspect vectors to guide relevance scoring and clustering. We obtain an optimum number of aspects from the aspect corpus and then reviewed the agenda of the conference in two different yeas, 2023 and 2024. Strategic terms are emphasized during weighting to enhance weak signal detection (Table 2). This process is shown in Appendix 9. The aspect-weighted vectors are later integrated with the initial topic model to build aspect-topic models, enabling alignment of new documents based on expert-defined themes. The list of aspects and their descriptions is presented in Appendix 3.

### *Keyword Definition and Weighting*

To evaluate our framework's responsiveness, we apply it to two cryptography-related aspects extracted from Quantum.Tech agendas across two consecutive years. These aspects capture shifts in cryptographic priorities, enabling iterative updates to the model. Table 2 presents the top TF-IDF-weighted keywords for each aspect.

**Table 2:** Two Aspect Keywords and Word Cloud: Weighted Keywords Across Protocols

| ***Word Cloud*** | **Top Keywords** |
|---|---|
| *Word Cloud of Aspect 1*<br>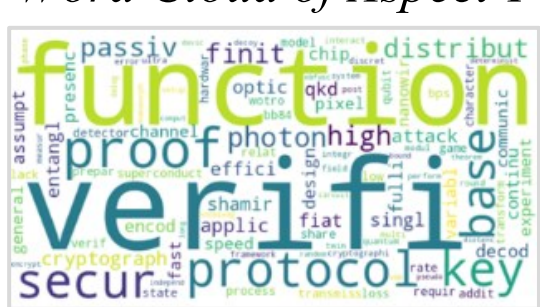 | verifi-0.029, function-0.022, proof-0.022, protocol-0.02, secur-0.019, key-0.019, base-0.019, photon-0.018, distribut-0.017, high-0.017 |
| *Word Cloud of Aspect 2*<br>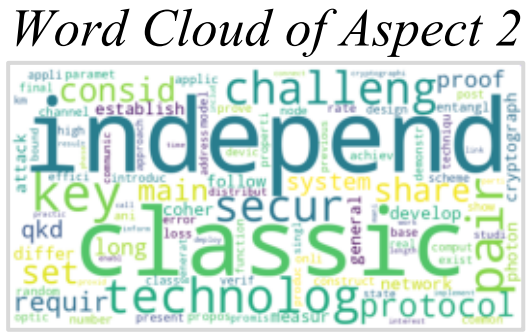 | independ-0.03, classic-0.026, key-0.025, pair-0.024, technolog-0.024, challeng-0.021, secur-0.02, protocol-0.02, share-0.02, set-0.02 |

These keyword sets reflect evolving focus areas, from verification and security protocols to independent architectures and implementation challenges. We evaluate the framework's ability to adapt to emergent terminology and capture directional shifts in expert discourse by comparing them over time.

Standard TF-IDF (Section 4.9) tends to overlook rare yet strategically important terms that signal emerging or disruptive developments. In order to address this shortcoming, we augment TF-IDF with expert-informed keywords, especially low-frequency terms flagged as critical by domain specialists, and assign them elevated weights. This integration improves the model's sensitivity to weak signals by boosting the relevance scores of documents containing such terms. When combined with topic distributions, this method surfaces emerging themes that traditional models may miss.

## 4.6. Phase 3: Align – Expert-Guided Alignment Layer

Each document was vectorized and evaluated for its similarity to each aspect using cosine similarity (Equation 2). Documents are then assigned to the aspect with the highest relevance score using a supervised clustering approach (Section 4.6.1). The effectiveness of aspect-based reassignment is illustrated through heatmaps of topic-document distributions (Figure 10). It shows clearer and more interpretable groupings in the aspect-guided model compared to the original LDA baseline.

### *Relevance Scores Computation*

To compute the relevance scores between weighted keywords and topic word distributions, combine the two-word distribution vectors using the formula below. This process quantifies the relationship between weighted aspect-stemmed terms (keywords) and their stemmed topic terms (keywords) across various aspect-topic models [1], [36]. Assume $A_i = \{a_{i1}, a_{i2}, \ldots, a_{in}\}$ be the set of keywords (i.e., *n*=50) for aspect *i* with weights $\mathrm{w}_{\mathrm{ij}}$, and $\mathrm{T_j} = \{\mathrm{t_{j1}}, \mathrm{t_{j2}}, \ldots, \mathrm{t_{jm}}\}$ be the set of keywords (i.e., *m*=100) for topic *j* with weights $\mathrm{v_{jk}}$. The relevance score ($\mathrm{R_{ij}}$) between aspect *i* and topic *j* is computed as

$$\mathrm{R_{ij}} = \sum_{\mathrm{k=1}}^{\mathrm{n}} \sum_{\mathrm{l=1}}^{\mathrm{m}} \mathrm{w_{ik}} \cdot \mathrm{v_{jl}} \cdot \mathrm{sim}\big(a_{\mathrm{ik}}, \mathrm{t_{jl}}\big) \text{ (2),}$$

where $\mathrm{sim}\big(a_{\mathrm{ik}}, \mathrm{t_{jl}}\big)$ is a similarity measure between the aspect keyword $a_{\mathrm{ik}}$ and the topic keyword $\mathrm{t_{jl}}$. This process helps us iteratively tune the topic model by weighting the keywords and incorporating external viewpoints outside the topic model as aspects of the domain. These scores show a strong link between the topic and aspect keywords. They reveal the relevance of the keywords to the main themes of the topics. There are several methods to find relevant scores. One uses cosine similarity between aspect terms and topic terms. This method measures the similarity of term vectors in both aspect and topic spaces. It gives a score that shows their relationship. Cosine similarity measures the closeness of the terms to the main topics. It does this by checking the angles between the vectors [36]. Our approach used the above formula (2) straightforwardly. It multiplies the weights of aspect terms with the corresponding topic terms (stems) in the topic model. The similarity measure equals 1 only when both term stems are the same. These scores help assess the relationship between topic words and aspect terms. They reveal how relevant the terms are to the main topics. In cryptography aspect, for instance, the term "challeng" has high relevance scores. They are 0.506 in topic T19 and 0.331 in topic T21. T19 is about security, computing, and cryptography. T21 is about cryptography, technology, and computing. This shows the Cryptography aspect's importance in both topics, as seen in Figure 8.

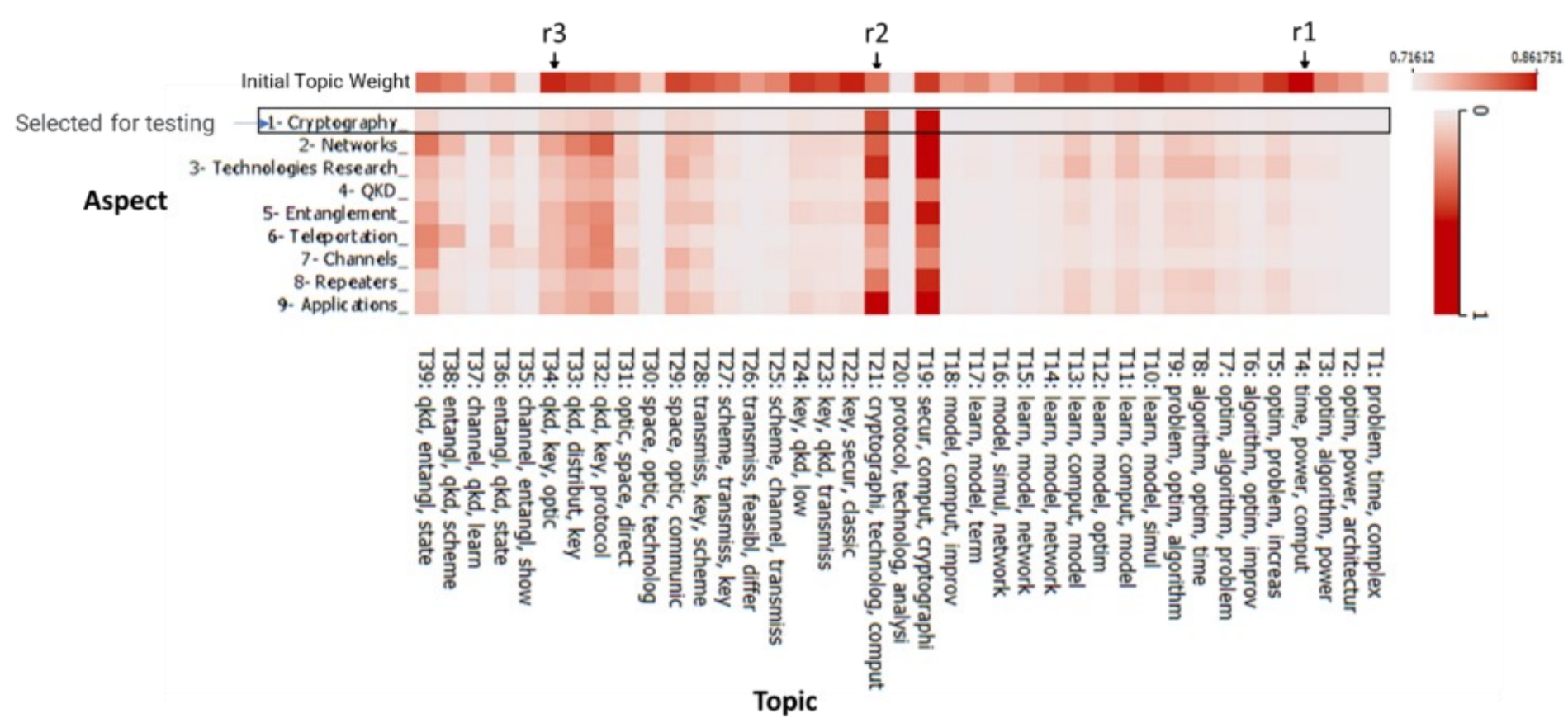

**Fig 8:** Aspects keywords, weighting and relevance scores with topics

Figure 8 displays the distribution of nine domain aspects across the 39 topics. The x-axis represents topics from the initial model (T1–T39), and the y-axis lists the aspects, including cryptography, entanglement, and teleportation. Each topic's association with aspects is visualized through the weighted presence of the top 50 aspect keywords. This heatmap enables quick assessment of how strongly each topic aligns with specific aspects, offering insight into thematic overlap and divergence. The underlying data, including the aspect-topic weights and top keywords, are available in the accompanying .xlsx file[6]. To illustrate the application, we constructed nine aspect-topic models using these keyword weights. As a case example, we focus on the cryptography aspect and analyze three representative cells (r1, r2, r3 topics), comparing their weights in both the initial and cryptography-specific topic models (Appendix 4). This reveals how topic-aspect alignment highlights emerging themes and supports deeper analysis in quantum communication research.

r1: This data point includes topic 4 documents with varied focus. Some are relevant to cryptography to a moderate extent (e.g., QKD, post-quantum cryptography). Others focus on optimization and routing, making them less relevant to cryptography. This is shown as cooler colors in the distributions (Figure 8).

r2: Topic 21's documents are very relevant to cryptography. They focus on post-quantum cryptography, secure communication, and quantum-resistant public key cryptography. The focus on cryptography in the documents raises their cryptography-related weights. The cell has a deeper color than the initial topic. So, the authors weighted the documents more with cryptography keywords.

r3: Topic 34 in the cryptography aspect model is about QKD and related quantum communication technologies. The focus on QKD, while relevant to secure communication, makes it less relevant to common cryptographic methods. This shows a niche in cryptographic research.

### *Applying a supervised clustering method*

Individual topics may hold important insights but may not be prominent in the initial model. To identify these insights, we apply a supervised clustering approach to assign new documents, DT and Appendix 6 to aspect-specific topics rather than relying solely on the baseline topic model. Following the method in [37], we use TF-IDF vectors to represent terms across 1 to *n* documents and select seed documents as initial cluster representatives. Each document is then assigned to the most similar aspect-topic using a similarity metric such as Euclidean distance, refined through thresholding. This supervised method builds on [38] to enhance weak signal detection. We evaluate cluster quality against labeled aspect-topics and track alignment over time. To test this, we also infer topics from the initial topic model and compare them with the aspect-topic assignments. Figure 9 presents an analysis of six sample documents, illustrating differences between the two models.

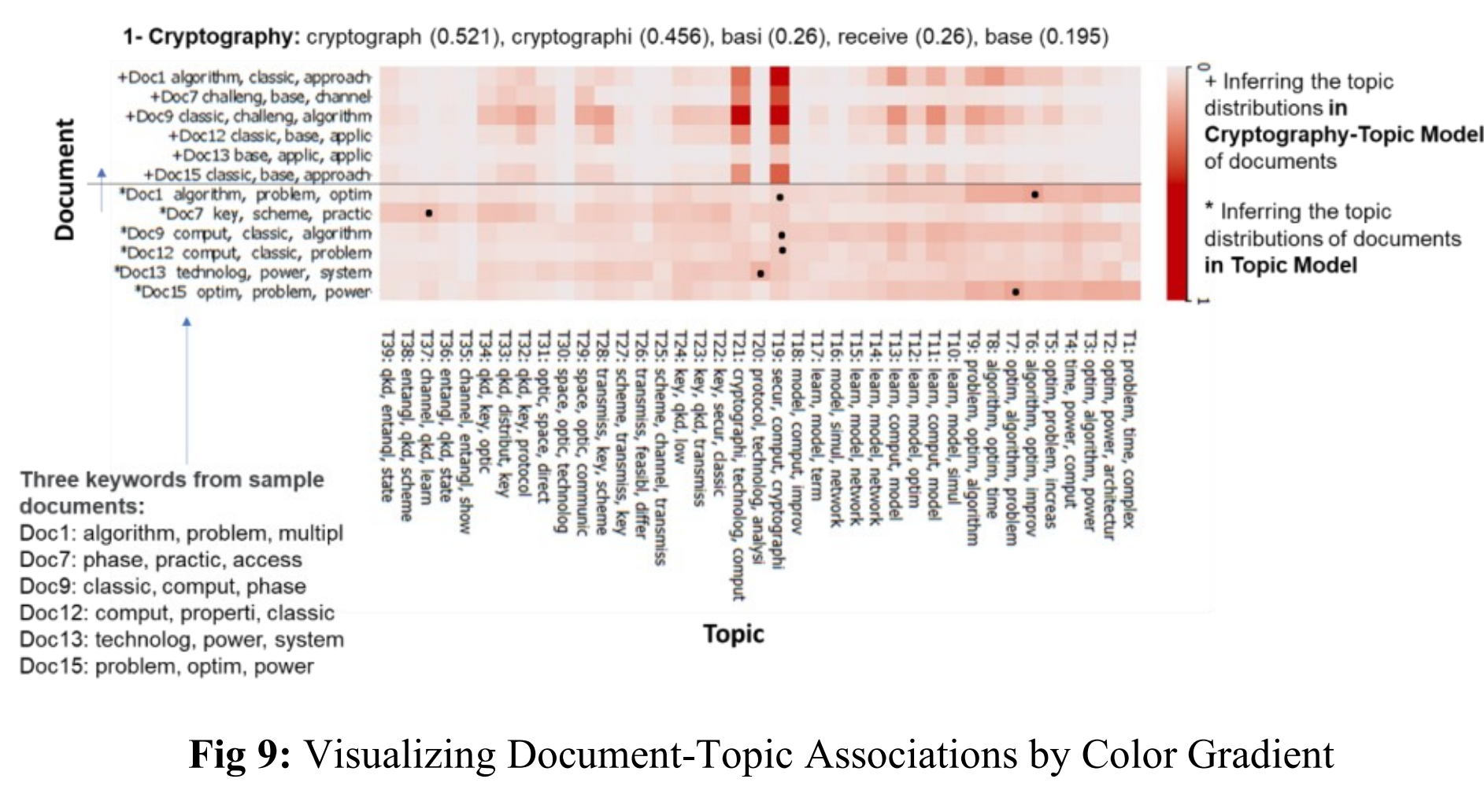

**Fig 9:** Visualizing Document-Topic Associations by Color Gradient

[6] https://github.com/AspectTopicModels

The x-axis represents 39 topics; the y-axis shows two document groups. The lower heatmap (asterisks) displays topic-document weights from the initial model, while the upper (plus signs) shows weights from the cryptography aspect model. Dotted cells in the lower heatmap highlight areas where the corresponding cell in the upper map has a contrasting shade. This contrast shows how aspect-based modeling reallocates documents to more thematically relevant topics. For example, while Document 9 is spread across many topics in the initial model, the cryptography model puts its weight on fewer, more relevant topics such as Topics 19 and 21. This indicates improved topic-document alignment. Overall, the aspect-informed model enables clearer clustering based on relevance that support more precise document analysis through supervised relevance scoring.

Appendix 5 lists the documents associated with the dotted cells in Figure 9. It highlights shifts in topic relevance between the initial and cryptography models:

Doc1 (algorithm, problem, multipl): It initially focused on quantum optimization (T6, 0.324), and with minimal cryptographic relevance (T6, 0.016). Its relevance to cryptography increases in T19 (from 0.17 to 0.203) that indicates partial overlap but non-central cryptographic focus.

Doc7 (phase, practic, access): This document initially assigned to T37 (0.078) and its weight drops to 0.017 in the cryptography model that confirm its limited cryptographic relevance.

Doc9 (classic, comput, phase): Weight of this document in T19 rises from 0.255 to 0.676 that shows a significant shift toward cryptographic relevance, possibly due to its application of quantum ML in secure systems.

Doc12 (comput, properti, classic): This weights in T19 and increases from 0.291 to 0.631, showing stronger alignment with cryptographic themes, particularly quantum vs. classical computing.

Doc13 (technolog, power, system): This document in T20 weight drops sharply from 0.124 to 0.003, revealing its focus on power systems is unrelated to cryptography.

Doc15 (problem, optim, power): T7 weight falls from 0.206 to 0.032, indicating that although it deals with quantum optimization, it lacks relevance to cryptographic methods.

After this illustration, we applied the two cryptography aspects from Table 2 in two separate iterations of our framework to this cryptography topic model as discussed. Each aspect, extracted from different time points, is used to guide document clustering and topic refinement. The impact of these iterations is demonstrated in the results section, showing how aspect evolution over time influences topic assignments and weak signal detection.

### 4.7. Phase 4: Stability and Evolution – Iterative Refinement

To evaluate the framework's robustness in detecting technological change, we implemented an iterative refinement process. After each round, aspect centroids are updated and documents reassessed using cosine similarity. Documents aligning more strongly with a new aspect-topic centroid are reassigned. This process continued until convergence when defined as fewer documents shifting between clusters across two consecutive iterations, and stabilization of within-cluster similarity scores. This iterative reassignment showed semantic cohesion and enhanced the clarity of emerging topic structures. It also confirmed that the model could maintain stable, interpretable clusters over time, a key capability for tracing weak signals and forecasting shifts in fast-moving domains like quantum cryptography. The stability of document-aspect alignments serves as a practical indicator of the framework's effectiveness in capturing evolving thematic relevance.

## 4.8. Alignment of Method Steps with Research Gaps

To improve traceability and clarify how our method addresses existing limitations, Table below provides a mapping between the steps of the proposed approach and the research gaps identified in the literature review. This alignment demonstrates how each procedural element, such as using TF-IDF to weight aspect keywords or clustering documents based on aspect-topic relevance, responds to known shortcomings in topic modeling, particularly the lack of expert integration, adaptability, and structured feedback loops.

**Table 3:** Alignment of Method Steps with Research Gaps

| Research Gap | Description | Method Steps Addressing the Gap |
|---|---|---|
| **Gap 1: Methodological** | Topic models lack contextual integration and rely on static configurations, limiting adaptability and relevance. | - Define initial corpus (Step 1) - Identify domain aspects (Step 4) - Extract weighted aspect keywords (Step 5) |
| **Gap 2: Practical** | Expert input is not dynamically or systematically integrated, reducing strategic impact. | - Build aspect corpora from expert sources (Step 4) - Compute similarity & supervised reassignment (Steps 6–7) - Create test corpus for domain adaptation (Step 2) |
| **Gap 3: Theoretical** | Few models support continuous learning or reward-driven mechanisms to optimize topic relevance over time. | - Iterative refinement loop (Step 8) - Document re-evaluation using aspect centroids (Step 8) - Evaluation using precision/coherence and expert review (Step 9) |

## 4.9. Formal Notation of the Proposed Method

To ensure reproducibility and theoretical clarity, this section presents the formal notation of the proposed Aspect-Weighted Topic Modeling method. Building on the earlier pseudocode and implementation steps, we define the core computational components of the model—including document representation, aspect vector construction, similarity scoring, and topic refinement, using structured mathematical definitions. This formalism provides a concise overview of the model's internal logic and enables future extensions, such as integration into reinforcement learning frameworks.

The initial corpus of domain-specific documents be denoted as C, where C = {$d_1$, $d_2$, ..., d_M}, with each $d_i$ representing a document and M being the total number of documents in the corpus. A subset of documents D_test from the original corpus C is selected as the test corpus to train the initial topic model (TM) generated by a topic modeling algorithm. The trained model outputs a set of topics, where each topic $T_j$ is represented as a list of keyword-probability pairs: $T_j$ = {($t_{j1}$, $v_{j1}$), ..., ($t_{jm}$, $v_{jm}$)}, with $t_{jk}$ denoting a keyword and $v_{jk}$ its corresponding weight or probability within topic $T_j$.

A set of expert-defined aspects is represented by A = {$A_1$, $A_2$, ..., $A_n$}, where each aspect $A_i$ refers to a high-level concept relevant to the domain (e.g., Cryptography, Quantum Key Distribution, etc.). Each aspect is associated with a raw text input, denoted as $AText_i$, which may include curated descriptions, summaries, or expert-defined definitions. The Quantum.Tech webcontent for expert-informed knowledge is used as a proxy to identify relevant aspects. Specifically, the 2022 Quantum.Tech report was used to extract nine core aspects, while the 2023 and 2024 cryptography reports were later used to define Aspect 1 and Aspect 2, respectively, for the two iterations of topic model refinement. For each aspect $A_i$, a weighted keyword vector is constructed by applying the TF-IDF algorithm to the corresponding $AText_i$. The resulting aspect vector is represented as: $A_i$ = {($a_{i1}$, $w_{i1}$), ..., ($a_{ik}$, $w_{ik})\}$, where $a_{il}$ is a salient keyword and $w_{il}$ is its TF-IDF score. The TF-IDF score for each term t in $AText_i$ is computed as:

$$\mathrm{TF-IDF}(t, AText_i) = tf(t, AText_i) \times \log(N / df(t)) \text{ (3)},$$

where tf(t, $AText_i$) is the frequency of term *t* in $AText_i$, df(t) is the number of aspect documents in which *t* appears, and N is the total number of aspect documents (i.e., N = |A|). While TF-IDF is a standard technique for keyword weighting, it is not the core contribution of this method. However, the novelty is in using TF-IDF to construct interpretable aspect vectors from expert-informed texts, and subsequently integrating these vectors into the topic modeling pipeline. This integration allows the model to align unsupervised topic outputs with strategic or domain-specific tasks where it facilitates more targeted topic interpretation and refinement. In this way, the method changes the focus from purely statistical topic discovery to a hybrid approach that incorporates human expertise and domain relevance insights. Each aspect vector $A_i$ and topic vector $T_j$ are then compared using a weighted cosine similarity function to measure alignment. The similarity score $S_{ij}$ between aspect $A_i$ and topic $T_j$ is computed as:

$$S_{ij} = (\Sigma\, w_i(t) \times v_j(t)) / (||A_i|| \times ||T_j||) \text{ (4)},$$

where the summation is taken over the shared vocabulary between $A_i$ and $T_j$, and the denominator denotes the product of the Euclidean norms of the respective vectors.

The result is a similarity matrix S, which maps each aspect to each topic and quantifies their alignment. These similarity scores can be used for several downstream purposes: identifying the most relevant topics for a given aspect, re-ranking topics for interpretability, or integrating the alignment scores as part of a reward signal in an extended reinforcement learning framework.

## 5. Results

To create two aspect topic modeling in two iterations, we applied two aspect 1 and 2 to cryptography topic model as first aspect topic model generated from applying LDA and then apply general cryptography aspect. These aspects represent two distinct cryptographic themes extracted from Quantum.Tech sources (see Table 2 and Figure 9). Each aspect-topic model is generated using its respective TF-IDF-weighted keyword vectors, capturing key concepts related to different subfields of cryptographic research.

Then, to evaluate the effectiveness of our aspect-guided topic modeling framework, we applied QCrypt 2023 and 2024 conference papers to Aspect 1- and Aspect 2-topic models to show how these new papers are align with the insight identified by applying two aspects in iterations results. We used QCrypt proceedings as external, time-stamped test corpora to assess the framework's ability to detect shifts in topical focus and semantic alignment. The test included QCrypt papers from each year, forming two evaluation sets. The process revealed clear distinctions between Aspect 1 and Aspect 2 models:

- Aspect 1 (e.g., proof-based cryptography, verification techniques) aligned more closely with theoretical papers from QCrypt 2023 papers focused on foundational protocols and security guarantees.
- Aspect 2 (e.g., device independence, implementation challenges) showed stronger relevance to QCrypt 2024 papers, which emphasized real-world deployment, classical compatibility, and operational feasibility.

The comparative word clouds (Table 6) demonstrate these alignments, where documents in each test set are scored against both models. A higher density of strong scores in the diagonal cells indicated a better alignment between QCrypt documents and their intended aspect-topic clusters. Besides, we observed that Aspect 1 topic model assigned higher weights to topics 34, 19, and 37 where areas focused on proof-based mechanisms and photon-level protocols (Appendix 7). But in second iteration and Aspect 2 emphasized topics 19, 32, and 39, reflecting interest in integration layers and hybrid systems with classical infrastructure (Appendix 8).

**Table 4:** Top Keywords of Selected Topics related to Aspect 1 in Iteration 1

| The selected topics | Topic Keywords |
|---|---|
| T34 | key(0.416), protocol(0.397), secur(0.385), distribut(0.371), optic(0.365), photon(0.360), channel(0.308), entangl(0.301), high(0.246), qkd(0.232) |
| T19 | secur(0.980), key(0.963), function(0.945), cryptographi(0.941), design(0.938), communic(0.938), applic(0.937), effici(0.919), base(0.872), protocol(0.868) |
| T37 | photon(0.015), qkd(0.014), key(0.011), state(0.010), protocol(0.010), distribut(0.009), experiment(0.008), effici(0.008), model(0.007), channel(0.002) |
| T32 | protocol(0.956), photon(0.950), secur(0.820), key(0.818), entangl(0.734), distribut(0.719), optic(0.602), rate(0.578), qkd(0.520), channel(0.499) |
| T24 | qkd(0.187), secur(0.170), distribut(0.160), key(0.145), rate(0.127), low(0.120), base(0.117), high(0.106), protocol(0.106), transmiss(0.095) |

In table 4 the most aligned topics, T19, T24, T32, T34, and T37 are listed based on their keyword relevance scores. These topics show strong alignment with foundational cryptography concepts of aspect 1 such as “key,” “protocol,” “security,” “QKD,” and “photon.” This reflects secure communication primitives central to quantum cryptographic discourse. For example, topic 19 emphasizes “security,” “key,” and “cryptography” with high weights (>0.9), marking it as the core semantic cluster for cryptographic function design and application. Besides, Topics 32 and 34 similarly highlight operational elements like “protocol,” “distribution,” and “entanglement,” consistent with QKD system architecture and photon-based key exchange. Also, Topic 24 has terms like “rate,” “transmission,” and “low,” indicating performance-oriented evaluations, while Topic 37 features experimental and modeling vocabulary, showing methodological diversity.

**Table 5:** Top Keywords of Selected Topics related to Aspect 2 in Iteration 2

| The selected topics | Topic Keywords |
|---|---|
| T19 | secur(0.981), technolog(0.977), key(0.971), challeng(0.954), comput(0.953), protocol(0.951), system(0.947), classic(0.942), develop(0.941), requir(0.937) |
| T32 | secur(0.713), qkd(0.632), protocol(0.612), key(0.551), photon(0.538), entangl(0.447), measur(0.437), channel(0.422), rate(0.410), classic(0.383) |
| T39 | entangl(0.467), photon(0.402), secur(0.390), key(0.358), qkd(0.341), channel(0.341), protocol(0.339), measur(0.329), state(0.322), scheme(0.266) |
| T21 | technolog(0.750), key(0.678), comput(0.637), cryptographi(0.630), protocol(0.603), classic(0.595), develop(0.595), secur(0.574), system(0.493), challeng(0.387) |
| T33 | secur(0.434), key(0.419), qkd(0.419), scheme(0.307), entangl(0.303), distribut(0.299), photon(0.293), protocol(0.292), measur(0.289), channel(0.280) |

To assess the framework captures emerging discourse around interoperability and post-quantum cryptography, we applied the second aspect to the topic model. Table 5 lists four dominant topics, T19, T21, T32, and T39, that showed the highest alignment with the Aspect 2. This aspect emphasized terms like “classical,” “technology,” “challenge,” and “system,” which reflect the transitionary challenges from classical to quantum-secured. Topic 19 remains central, though its focus evolves from foundational security (as seen in Aspect 1) to emphasize terms like “technology,” “challenge,” and “requirement.” It is revealing a shift toward integration and system-level concerns. Topic 21 is highly weighted for “technology,” “develop,” “comput,” and “classic,” that directly reflecting post-quantum strategy formulation and classical system compatibility, core themes in QCrypt 2024 papers. Topics 32 and 39 also show significant alignment. Topic 32 bridges operational QKD elements (“rate,” “channel,” “classic”), and Topic 39 focuses more on technical enablers such as “entanglement,” “measurement,” and “scheme.” These alignments confirm the model's sensitivity to evolving cryptographic architectures and the emergence of hybrid classical-quantum communication strategies.

To validate our framework’s sensitivity to evolving discourse, we analyzed QCrypt papers from 2023 and 2024 across both aspect-topic modeling iterations. Table 6 compares the alignment between selected topics and their corresponding high-ranking documents from each year. This two-year comparison confirms that topics identified as relevant in the initial model (Aspect 1) continue to evolve in content and emphasis through the second iteration (Aspect 2).

**Table 6:** Cryptography Aspect Performance Across Iterations

| Topic | First Iteration Topic Model Words | Top Docs from QCrypt 2023 in iteration 1 | Second Iteration Topic Model Words | Top Docs from QCrypt 2024 in iteration 1 |
|---|---|---|---|---|
| T19 | 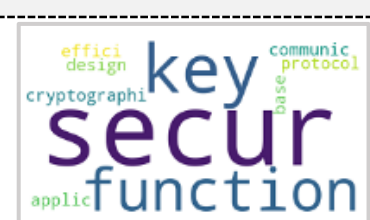  | 1 -18-8 -32-3 | 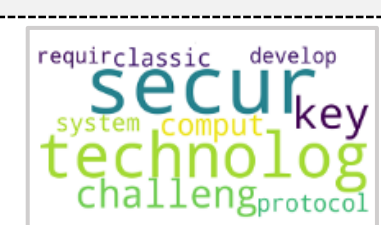  | 22-16-21-12-6 |
| T32 | 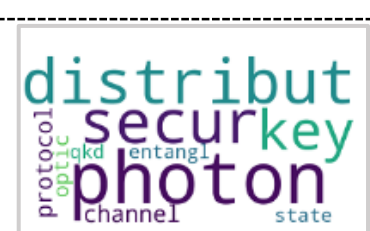  | 32-12-10-33-2 | 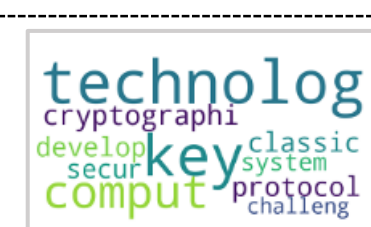  | 16-22-11-12-21 |
| T39 | 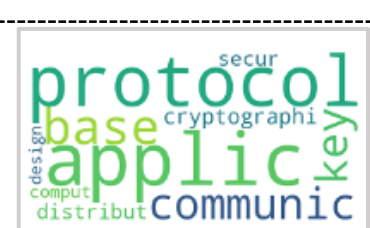  | 32-10-12-33-8 | 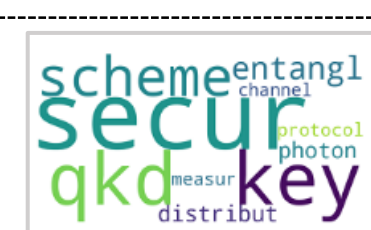  | 16-11-12-22-31 |
| T21 | 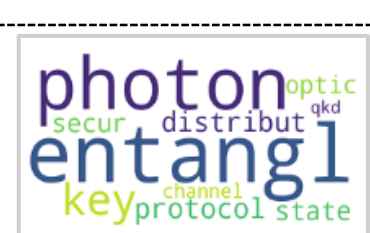  | 18-1 -32-8 -9 | 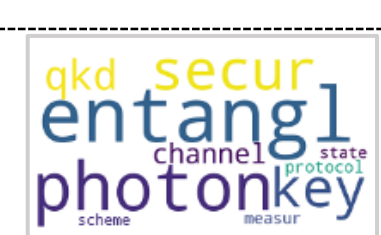  | 16-11-22-12-21 |
| T33 | 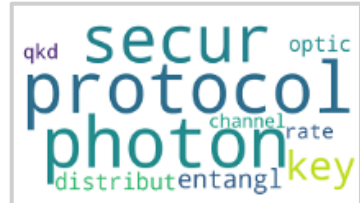 | 10-32-12-33-8 | 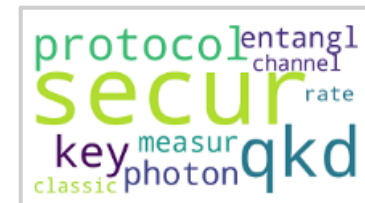 | 22-16-6 -31-5 |

The word clouds and topic-keyword distributions reveal that documents from QCrypt 2023 are strongly aligned with foundational cryptographic concerns, such as “key,” “protocol,” “photon,” and “distribution”, dominant in topics T19 and T32. These topics are the core infrastructure and security primitives of QKD systems at the time. In contrast, the QCrypt 2024 documents show a clear shift in focus, reflected in topics such as T21 and T39, which emphasize “technology,” “challenge,” “classic,” and “system.” These keywords show a growing interest in interoperability, post-quantum transition strategies, and integration with classical cryptographic infrastructure.

The evolving activation of topics across years, highlighted by the shifting document-to-topic assignments in Table 6, demonstrates that our iterative, aspect-weighted modeling approach can detect and track changes in technological discourse. This confirms the utility of the framework not only for topic enhancement and document alignment, but also for forecasting trends and identifying early signals of strategic importance in fast-moving domains like quantum cryptography.

## 6. Evaluation of the Framework

The proposed framework is evaluated on three key dimensions. The dimensions are topic coherence, alignment with real-world expert themes (derived from QCrypt 2023–2024 papers), and topic stability over iterations. The results show that the aspect-guided models weighted more coherent and interpretable topics than the baseline model. The keywords in Tables 5 and 6 reflect that how topics focused on cryptographic themes across iterations. For example, topics T19, T32, and T21 prominently weighted terms like key, secur, protocol, and qkd that are aligned with the aspects. Besides, to validate alignment with expert-informed themes, we refined the models to test documents from QCrypt 2023 and 2024. Table 6 shows how topic words have changed and how these documents are repeatedly assigned to the same high-relevance topics across both modeling iterations, confirming consistency and relevance. Notably, documents such as D12, D16, and D22 were assigned to T32 and T19 in both aspect-topic models, showing strong thematic consistency with the cryptographic keywords. This supports the model’s ability to detect and track emerging signals and trends over time. Furthermore, the topic stability improved between iterations such as in T19, as shown by the reduced reassignment rate (28% to 4.2%) and increased intra-cluster similarity (0.61 to 0.74). In sum, these results confirm that the framework not only enhances topic interpretability but also aligns effectively with expert-informed knowledge and sustains stability during refinement.

### 6.1. Comparison with Prior Work

Recent work in topic modeling has expanded into various directions, including semi-supervised seeding, domain adaptation, external knowledge use, and, more recently, reinforcement learning. However, very few methods combine these ideas into an integrated framework suitable for real-world strategic applications such as technology mapping.

The table below compares our method with few influential or recent studies from 2012 to 2024, highlighting the key features of our work. Our framework stands out for its combination of weighted aspect keyword modeling, document inference through topic-aspect alignment, and support for iterative topic refinement. Most models (e.g., [19], [39] rely on static seed words and do not include document reassignment or feedback. Others (e.g., [26] explore reinforcement learning, but without domain-guided external knowledge or structured aspect modeling.

**Table 7:** Comparative Overview of Recent Studies on Aspect-Based Topic Modeling

| Study | Aspect Integration | Dynamic Weighting | Doc Inference or Reassignment | Iterative Refinement | External Knowledge | Domain Focus / Application |
|---|---|---|---|---|---|---|
| Jagarlamudi et al. (2012) | Seed terms | Static | | | Manual seeds | General NLP |
| Nguyen et al. (2015) | Seeded, embeddings | Indirect via vectors | | | Pretrained embeddings | General |
| Seo et al. (2023) | | Learned relevance | ✓ RL-driven adaptation | ✓ RL topic | Internal only | AI language agents |
| Coccia & Roshani (2024) | ✓ Expert topic framing | Static clusters | | | ✓ Strategic foresight | Technology monitoring |
| Watanabe & Baturo (2024) | ✓ Seeded sequential LDA | ✓ Semi-supervised | ✓Sentence-level classify. | | ✓ Seed words | Social science |
| Zhang et al. (2021) | ✓ External vocab + co-occurrence | ✓ Weighted co-matrix; | Clustering step | | ✓ Corpus + lexicon | Health informatics |
| **This paper (Proposed Method)** | ✓ External aspect corpora | ✓ TF-IDF weighting; similarity | ✓ Inference-based reassignment; | ✓ Iterative, RL-compatible | ✓ Expert-based | Quantum technology |

Studies like [4] incorporate expert framing but do not provide a formal topic modeling pipeline or [40] that extended LDA with a semi-supervised topic modeling using seed words without iterative refinement. Models like [5] incorporate aspects, but use internal weighting rather than corpus-derived TF-IDF with cosine-based scoring. Our work is distinct in its multi-stage inference framework, use of external structured texts, and design for adaptability, making it suitable for identifying weak signals in fast-evolving domains like quantum cryptography.

### 6.2. Validation Approach

We followed an internal validation to ensure the artifact is built correctly, aligned with the Design Science Research Methodology (DSRM) outlined in[31], and suitable for its intended use in technology monitoring within the quantum communication domain. Validation process focused on three key dimensions:

• Design Traceability and DSRM Alignment

The framework was developed following the Design Science Research Methodology (DSRM), as described in Chapter 3. Components including corpus preparation, aspect definition, topic modeling, alignment, and refinement, are explicitly mapped to the corresponding DSRM phases, including problem identification, artifact design, demonstration, and evaluation. This process ensured logical iteration, theoretical grounding, and alignment with the design requirements identified in Chapter 3.

• Use Case and Expert-Proxy Alignment

The framework was validated against its intended application that supports foresight and thematic structuring in the quantum communication domain. Selected external sources, QCrypt 2023 and 2024 conference proceedings, were used as expert-informed proxies for validate emerging technologies from the models. These sources helped define the aspects and guided the alignment logic, ensuring that the method reflected real-world subthemes relevant to early signal detection and domain-specific mapping.

• Internal Logical Coherence and Modularity

The framework's modular structure allows repeatability and adaptation across different domains. Each phase builds systematically on previous outputs, avoiding ad hoc decisions. The use of formal notation in Sections 4.6.1 and 4.9 further demonstrates the method's computational transparency and supports its reproducibility and future integration with learning-based refinement (e.g., reinforcement learning environments).

Together, these validation activities confirmed that the framework was developed in a systematic, expert-informed, and purpose-driven manner, providing a solid foundation for the performance evaluation presented in the subsequent sections.

## 7. Discussion

Our study's results show how to adapt document classifications when adding aspects to topic models. We used a comparative analysis of topic and aspect models. It evaluated the thematic alignment and significance of documents within different aspects. This test lets us measure how well keywords improved topic relevance. It also revealed a shift in focus towards a specific research aspect. This section interprets the evaluation results and reflects on the implications of the proposed method for refining topic models through expert-informed aspect weighting. This section explores how well the framework addresses real-world information needs, compares with existing approaches, and what future directions emerge from its current design.

### 7.1. Insights from Evaluation

The evaluation shows that the method effectively realigns topic models with emerging expert knowledge in cryptography. Across two iterations we had cryptography-relevant topics (T19, T21, T32) that were consistently reinforced through weighted aspects derived from Quantum Tech conference materials. Besides, semantic alignment with benchmark QCrypt papers improved, confirming the method's ability to detect meaningful shifts in discourse. Furthermore, heatmaps and document-topic associations (e.g., Figure 9) demonstrated clearer topical clustering after each refinement. The iterative aspect injection helped maintain topic relevance over time while absorbing newer terms such as entanglement, network integration, and classical-quantum merging in the second round. This confirms that domain-grounded, externally informed topic adjustment can produce more relevant, interpretable, and up-to-date topic structures, without retraining the model from scratch.

### 7.2. Answers to the Research Question

This study addressed the main research question that is how can external expert knowledge be systematically integrated into topic modeling to improve contextual alignment and adaptability in specialized domains?

The proposed framework shows that integrating aspect keywords, selected from authoritative sources like a conference proceeding, provides targeted refinement of topic models. The framework reweighted topic-word distributions by weighting and embedding TF-IDF-ranked aspect terms into the modeling process to reflect strategically significant concepts such as QKD protocols, entanglement, and post-quantum security. The refined topics (e.g., T19, T21, T32) aligned more closely with expert-informed discourse by empirical evaluation with QCrypt papers. The method's iterative structure helped it adapt topic representations over time. For example, adding new aspect signals from 2024 created distinct cryptographic subthemes. These included classical-quantum interoperability and implementation challenges, all without needing to retrain the entire model. These results show that expert-informed aspect weighting is a scalable and effective way to enhance topic relevance and adaptability in rapidly changing technical fields.

### 7.3. Strategic Implications

The results show important implications for research and practice:

Real-World Alignment: Using keywords from trusted conferences keeps the model in line with current trends in cryptography. This is vital for staying ahead in fast-changing areas.

Efficient Updating: The framework allows for focused topic updates without needing to retrain on the whole dataset. This makes it ideal for making it well-suited for dynamic, time-sensitive knowledge tracking.

Interpretability: Unlike black-box methods, each step of refinement is clear and understandable. It includes weighted keywords, alignment heatmaps, and relevance scores.

Domain Adaptability: This method works well in quantum cryptography but can also apply to other fields with structured expert-informed input. This makes it useful across various specialized areas.

## 7.4. Comparative Positioning

The aspect-guided method demonstrates advantages in generating highly relevant, semantically coherent, and domain-aligned topics, particularly within cryptography-focused corpora in compared to existing alternatives. Table 8 provides a comparative analysis of common topic modeling and knowledge extraction approaches, highlighting their respective strengths and limitations.

**Table 8:** Comparative Analysis of Topic Modeling and Knowledge Extraction Methods

| Method | Advantages | Limitations |
|---|---|---|
| Baseline LDA | Fully unsupervised, fast | Ignores domain semantics; diluted topics |
| LLMs summarizers | Language fluency; flexible input | Opaque structure; unclear alignment tracking |
| Keyword Curation | High specificity | Not scalable; subjective |
| Our Proposed Method | Transparent, expert-informed, aligned with real-world trends | Requires optimization decisions techniques |

This positions the proposed framework as a balanced solution, more adaptive and domain-sensitive than unsupervised models, while remaining more transparent and controllable than black-box LLM-based summarization methods.

## 7.5. Limitations

Several limitations exist in our framework, despite promising results:

First, there is no optimization-based topic selection. We chose five main topics represent the cryptography aspect, but selecting these topics was made manually after analyzing topic distributions and associated documents. The process needs an optimization mechanism to rank or select the most relevant topics automatically. Future work could add criteria like coherence-maximizing or information gain to automate this decision.

Second, TF-IDF weighting helped extract aspect keywords. However, constructing and validating aspect corpora still depends on expert judgment. This reliance may introduce subjectivity and limit scalability. Semi-automated methods for curating corpora from structured metadata or taxonomies could ease this dependency.

Finally, the framework uses cosine similarity to compare document vectors with aspect vectors and assess cluster cohesion. While effective, cosine similarity might miss deeper semantic relationships. Future improvements could explore alternatives like phrase-based embedding distances, semantic entailment models (e.g., SBERT), or topic-aware alignment functions to better capture context and meaning.

These limitations suggest promising ways to enhance the framework's robustness, interpretability, and automation in future iterations.

### 7.6. Future Directions

Future enhancements to the framework will improve automation, scalability, and generalizability. Dynamic topic ranking could be optimized using decision techniques like reinforcement learning (RL). By applying scoring functions such as novelty or burstiness, the model can prioritize topics in real time. This approach does not depend on external annotations. The framework will benefit from cross-domain evaluation. It will test adaptability and robustness in high-stakes areas like AI policy, synthetic biology, and cybersecurity. Finally, incorporating weak labels from structured metadata could help build semi-supervised evaluation datasets. This includes information like author affiliations, session names, or topical tags. It will also enable the use of comparative metrics such as topic purity, recall, and coverage.

## 8. Conclusion

This study presents an aspect-guided method to refine topic models using weighted expert-informed keywords. It focuses on enhancing topical signals in emerging fields like quantum cryptography. The method boosts alignment by adding curated aspect terms in a two-phase refinement process. This is based on real-world expert feedback from Quantum Tech conferences and validated with QCrypt 2023 and 2024 publications. In both evaluation phases, the method improved document-topic alignment, topic interpretability, and the relevance of output topics to current cryptography discussions. Topics T19 and T21 are strengthened in the first iteration using cryptography Aspect 1 from Quantum Tech 2023. The topics aligned well with QCrypt 2023 papers such as key documents like Doc 9 and Doc 3. Aspect 2 from Quantum Tech 2024 refined the topic structure further. Topic T32 gained prominence, highlighting new research on entanglement and system integration. The relevance scores between aspect keywords and topic terms reached 0.506 for T19 and 0.331 for T21, confirming that aspect-weighting sharpened focus. Visual heatmaps showed clearer clustering of cryptography-related documents and less semantic drift across iterations. This framework stands apart from traditional models like LDA and black-box LLM summarizers. It offers transparency, modularity, and domain responsiveness without needing complete model retraining. It strikes a balance between semantic consistency and adaptability, which is perfect for tracking rapidly changing research areas. The results reveal that expert-informed aspect weighting is a simple yet effective way to boost the relevance and clarity of topic modeling outputs. Applying it iteratively helps align the model with shifting domain signals.

## Abbreviations

| | |
|---|---|
| BERT | Bidirectional encoder representations from transformers |
| C-V | C-V Coherence Value (a measure of topic quality in topic models, like LDA) |
| DSRM | Design Science Research Methodology |
| LDA | Latent Dirichlet allocation |
| LLM | Large Language Model |
| PQC | Post-Quantum Cryptography |
| QKD | Quantum Key Distribution |
| Regexp | Regular expression |
| RL | Reinforcement Learning |
| SBERT | Sentence-BERT |
| TF-IDF | Term frequency-inverse document frequency |
| UMass | University of Massachusetts Coherence Score (Topic Coherence Metric) |

## Funding

This work is not supported by any external funding.

### Authors' contributions

AN conceived the study, carried out the investigation, and developed the method. He wrote the original draft and contributed to the writing and editing of the manuscript. He also did a formal analysis, oversaw the project, validated the findings, and made visualizations. MW contributed by analyzing data, providing methodological expertise, and supervising the research. He also helped review and edit the manuscript. Both authors read and approved the final manuscript.

### Availability of data and materials

The datasets generated and/or analyzed during the current study are available in the GitHub repository, https://github.com/alinazari1/FineTuning/blob/main/

### Competing interests

The authors declare that they have no competing interests

### Ethics approval and Consent to Participate

Not applicable

### Consent for publication

All authors have reviewed and approved the final manuscript for publication. They consent to the publication of this work in its current form.

## Appendix

**Appendix 1:** The articles with the highest relevance to the review questions

*The themes include S1-Topic Modeling, S2-Topic Examination, S3-Multi-faceted Insights of a Topic S4-Multi-faceted Insights Supervised (or Semi) Clustering, and S5-Reinforcement or Deep Learning as depicted in Figure 1.

| Paper | S1 | S2 | S3 | S4 | S5 | Relevance |
|---|---|---|---|---|---|---|
| [8] | ✓ | | ✓ | | | Clustering Massive-Categories and Complex Documents via Graph Convolutional Network. |
| [41] | | | ✓ | | | Semi supervised Fuzzy Clustering with Partition Information of Subsets. |
| [22] | | | ✓ | ✓ | | Research proposal content extraction using natural language processing and semi-supervised clustering: A demonstration and comparative analysis. |
| [21] | | | ✓ | ✓ | | Document clustering with dual supervision through feature reweighting. |
| [28] | | | ✓ | | | Intention-guided deep semi-supervised document clustering via metric learning. |
| [23] | ✓ | | | | | Automatic constraints generation for semisupervised clustering: experiences with documents classification. |
| [42] | ✓ | ✓ | | | | Using structural topic modeling to identify latent topics and trends in aviation incident reports. |
| [43] | ✓ | ✓ | ✓ | | | Dataless Text Classification: A Topic Modeling Approach with Document Manifold. |
| [44] | ✓ | | ✓ | ✓ | | Self-organizing weighted incremental probabilistic latent semantic analysis |
| [45] | ✓ | | ✓ | | | Research on Multi-label Text Classification Method Based on tALBERT-CNN. |
| [46] | ✓ | | ✓ | ✓ | | tBERT model for semantic similarity detection |
| [47] | ✓ | | ✓ | | | Overview of topic modeling in bioinformatics. |
| [48] | | | ✓ | ✓ | | Supervised labeled latent Dirichlet allocation for document categorization. |
| [49] | ✓ | | ✓ | | | Supervised topic model for document classification |
| [18] | | | | ✓ | ✓ | Selective supervised Latent Dirichlet Allocation for document classification |
| [50] | ✓ | | ✓ | ✓ | ✓ | Regularized supervised topic model for multi-class regression |
| [51] | | | | ✓ | ✓ | Labelset topic model for multi-label document classification |
| [48] | | | | ✓ | ✓ | Centroid prior topic model for multi-label classification |
| [52] | | | | ✓ | | Effective document labeling with very few seed words |
| [53] | | | | ✓ | | Multi-label dataless text classification with topic modeling |
| [54] | | | | ✓ | | Seed-Guided Topic Model for Document Filtering and Classification |
| [55] | ✓ | | | | | Community Detection in Social Networks Considering Topic Correlations |
| [56] | | | | ✓ | ✓ | Seed-guided deep document clustering |
| [57] | ✓ | ✓ | | | | Topic extraction from short texts with regularization |
| [58] | ✓ | ✓ | | | | Community detection in social networks considering behaviors |
| [59] | ✓ | | | ✓ | | Neural topic model with multi-level supervisions |
| [60] | ✓ | | | ✓ | ✓ | Hierarchical classification of GitHub repositories using keywords |
| [61] | ✓ | | | | | Classification of texts into categories of cybersecurity incidents with topic models |
| [62] | | ✓ | ✓ | | | Revealing latent topical facets and their associated sentiments within weblog collections. |
| [63] | ✓ | ✓ | ✓ | | | Identifying topics and their underlying aspects, providing improved topic coherence and capturing multiple dimensions within a corpus |

## Appendix 2

**Q1.1:** How do classification or clustering methods contribute to refining topics through the use of weighted keywords? This question addresses the methods by which classification or clustering algorithms utilize weighted keywords to refine the identification and categorization of topics within a dataset. It involves techniques such as feature extraction TF-IDF (Term Frequency-Inverse Document Frequency) weighting and other methods aimed at assigning importance or relevance to specific keywords within documents. The goal is to enhance the precision and accuracy of topic identification and clustering by giving more weight to keywords that are deemed more significant or representative of the underlying themes.

**Q1.2:** What are the strengths and limitations of using these methods compared to our research method? This question focuses on evaluating the advantages and drawbacks of employing classification or clustering methods particularly in comparison to a specific research methodology. It involves assessing factors such as computational efficiency scalability interpretability robustness to noise or outliers' ability to handle high-dimensional data and suitability for the specific research context. Strengths may include improved automation scalability to large datasets and potential for uncovering hidden patterns or relationships. Limitations may encompass challenges related to algorithmic complexity sensitivity to parameter settings and dependence on labeled data for supervised methods.

**Q2.1:** The role of supervised clustering methods in discerning nuanced aspects within topics. This question delves into the significance of supervised clustering techniques in identifying subtle variations or nuances within overarching topics or themes. It involves exploring how supervised methods leverage labeled data or feedback to guide the clustering process towards more granular distinctions within clusters. Such techniques may involve the incorporation of domain knowledge expert feedback or human annotations to refine the clustering results and extract deeper insights from the data.

**Q2.2:** How do these techniques augment the interpretation and comprehension of clustered documents by revealing latent insights and subtle perspectives? This question addresses the impact of supervised clustering methods on enhancing the interpretability and comprehension of clustered documents by uncovering latent insights and subtle perspectives. It involves discussing how these techniques facilitate the extraction of meaningful patterns trends or relationships from complex data thereby aiding in the identification of underlying themes sentiment analysis trend detection and other forms of data-driven analysis. By providing more nuanced clustering results these methods contribute to a richer understanding of the underlying data structure and support informed decision-making processes.

**Appendix 3:** Terms of aspects of Quantum Communication Topics

| Aspect | High Frequent Terms (TF-IDF Scores) | Description |
|---|---|---|
| Cryptography | cryptograph (0.521), cryptographi (0.456), basi (0.26), receive (0.26), base (0.195) | Utilizes principles of quantum mechanics for secure communication, employing Quantum Key Distribution (QKD) to exchange unbreakable cryptographic keys between parties, ensuring unconditional security based on quantum laws. |
| Networks | classic (0.393), distribut (0.393), entangl (0.393), enabl (0.262), applic (0.197) | Establishes quantum communication infrastructure leveraging quantum entanglement, enabling secure transmission of quantum information over large distances, crucial for secure communication, distributed quantum computing, and quantum-enhanced sensing applications. |
| Technologies Research | develop (0.361), classic (0.289), effici (0.289), aim (0.217), algorithm (0.217) | Advances capabilities and applications of quantum communication systems through multidisciplinary efforts, including developing protocols, hardware, algorithms, and integration strategies, aiming for secure, efficient, and reliable transmission of information. |
| Quantum Key Distribution - QKD | send (0.32), receive (0.32), detect (0.32), applic (0.24), attempt (0.24) | Pioneering technology in quantum communication, offers unconditional security guarantees by exploiting quantum properties to exchange cryptographic keys between parties, ensuring interception detection through the fundamental principles of quantum mechanics. |
| Entanglement | distribut (0.398), enabl (0.318), applic (0.239), capabl (0.239), channel (0.239) | Entanglement is a profound quantum phenomenon where the states of two or more particles become inherently correlated, enabling revolutionary applications such as Quantum Key Distribution (QKD), quantum teleportation, and secure communication. |
| Teleportation | entangl (0.553), send (0.387), receive (0.332), challeng (0.221), distant (0.221) | Quantum teleportation allows for the instantaneous transfer of quantum information from one location to another, leveraging entanglement to transmit quantum states without physically moving particles, with applications in secure communication and distributed computing. |
| Channels | channel (0.923), challeng (0.185), applic (0.148), classic (0.111), atmospher (0.074) | Quantum channels are essential pathways for transmitting quantum information, such as photons or qubits, over various mediums like optical fibers or free-space links, necessitating techniques for maintaining coherence and fidelity to enable secure and efficient communication. |
| Repeaters | distanc (0.575), challeng (0.288), distribut (0.288), enabl (0.288), correct (0.23) | Quantum repeaters extend the reach of quantum communication by mitigating signal loss and decoherence over long distances, leveraging entanglement swapping and quantum memories to regenerate entangled states and ensure the reliability of quantum communication networks. |
| Applications | cryptographi (0.617), applic (0.393), develop (0.28), cryptograph (0.224), distribut (0.224) | Applications of quantum communication span various domains, including secure communication, quantum computing, and quantum sensing, offering transformative solutions such as Quantum Key Distribution (QKD), quantum teleportation, and quantum-resistant cryptography. |

**Appendix 4:** Initial vs. Cryptography Topic Models

| Doc ID - Weights in Initial – Cryptography Topic Models | Documents in Initial Topic Model | Documents in Cryptography Topic Model |
|---|---|---|
| r1-Topic 4: Less Relevance to Cryptography | | |
| 142 – (0.87 - 0.08) | Explores quantum computing for optimizing resource assignment in network management. | Focused on resource optimization, not cryptographic techniques. |
| 181 – (0.84 - 0.06) | Proposes AI-based energy-efficient routing for IoV using Quantum Chemical Reaction Optimization. | Focused on energy-efficient routing, not cryptographic methods. |
| 208 – (0.94 - 0.12) | Introduces a quantum game theory-based recovery model for fault tolerance in mobile computing. | Focused on fault tolerance, not cryptographic security. |
| 352 – (0.85 - 0.04) | Studies satellite-based QKD for global quantum internet, optimizing key distribution. | Related to cryptography via satellite-based QKD. |
| 472 – (0.87 - 0.05) | Investigates post-quantum cryptographic implementations for IoT, improving performance and energy efficiency. | Related to cryptography, focusing on post-quantum implementations. |
| 740 – (0.83 - 0.11) | Develops a high-performance FFT for space-borne radar altimeters using DSP technology. | Focused on FFT implementations, not cryptographic methods. |
| 748 – (0.93 - 0.31) | Optimizes lattice-based cryptographic algorithms for embedded systems. | Related to cryptography, focusing on post-quantum algorithms. |
| 978 – (0.84 - 0.05) | Proposes efficient cryptographic solutions for IoT and smart cities, enhancing data security. | Related to cryptography, focusing on post-quantum environments. |
| r2-Topic 21: High Relevance to Cryptography | | |
| 960 – (0.37 - 0.99) | Explores post-quantum cryptographic security for DTLS using PQ TESLA, securing 5G network messages. | Uses post-quantum TESLA to secure DTLS. |
| 762 – (0.28 - 0.99) | Discusses multivariate polynomials in PKC, focusing on the MQ problem, promoting post-quantum cryptography. | Discusses MQ problem-based PKC, resistant to quantum attacks. |
| 795 – (0.32 - 0.98) | Introduces NetQASM, a platform-independent instruction set for quantum internet applications. | Supports cryptographic applications in quantum internet. |
| 652 – (0.33 - 0.98) | Reviews structured light for high-capacity and high-security communication. | Explores structured light for high-security communication. |
| 233 – (0.50 - 0.93) | Implements a 24-hour bit commitment protocol for secure digital signatures and voting. | Ensures secure digital signatures and voting. |
| r3- Topic 34: Specific Focus on QKD and Quantum Communication – Less relevance | | |
| 29 – (0.70 - 0.05) | Proposes a fully connected QKD network using wavelength and space division multiplexing. | Focused on QKD network for entanglement distribution. |
| 51 – (0.80 - 0.04) | Describes a hand-held free-space QKD system using the BB84 protocol. | Focused on free-space QKD for short-distance communication. |
| 56 – (0.82 - 0.06) | Presents an experimental DIQKD system using entanglement between rubidium atoms for secure key exchange. | Focused on DIQKD for secure key exchange. |
| 66 – (0.80 - 0.03) | Theoretically investigates optical systems for quantum computation, proposing a superstructure for quantum sampling. | Focused on quantum computation, not directly on cryptography. |
| 92 – (0.79 - 0.06) | Reports on a compact quantum dot single-photon source for QKD at telecom wavelengths. | Focused on single-photon source for QKD in fiber-based networks. |

**Appendix 5:** Initial vs. Cryptography Topic Models with test documents

| Test Doc ID/ Terms | Document Weight in Initial - Cryptography Topic Models | Documents in Initial Topic Model | Documents in Cryptography Topic Model |
|---|---|---|---|
| Doc1: algorithm, problem, multipl | T6, T19: 0.324, 0.17 - 0.016, 0.203 | Quantum algorithm for higher-order unconstrained binary optimization and MIMO maximum likelihood detection. | Not focused on cryptographic methods or secure communication. |
| Doc7: phase, practic, access | T37: 0.078 - 0.017 | Experimental quantum secret sharing based on phase encoding of coherent states. | Relevant to cryptography as it deals with quantum secret sharing. |
| Doc9: classic, comput, phase | T19: 0.255 - 0.676 | Quantum machine-learning phase prediction of high-entropy alloys. | Not focused on cryptographic methods or secure communication. |
| Doc12: comput, properti, classic | T19: 0.291 - 0.631 | Quantum vs classical computing: a comparative analysis. | Not specifically focused on cryptographic methods. |
| Doc13: technolog, power, system | T20: 0.124 - 0.003 | Disruptive technology identification method for new power systems based on patent evolution analysis. | Not focused on cryptographic methods or secure communication. |
| Doc15: problem, optim, power | T7: 0.206 - 0.032 | Annealing-based quantum computing for combinatorial optimal power flow. | Not focused on cryptographic methods or secure communication. |

**Appendix 6:** Pseudocode of the Proposed Method

*Phase 1: Explore – Topic Detection*

1: SK <- define_search_keywords(domain)
- The domain of interest is specified.
- Search keywords are derived from key papers and relevant documents.
- Inclusion and exclusion criteria are applied to ensure specificity and relevance.

2: DL <- get_domain_specific_corpora(SK)
- Collect various document types: peer-reviewed articles, book chapters, surveys, reviews, conference papers.
- Sources: Two online libraries, Web of Science and Scopus.
- Expand collection using citations and Litmaps tool for visualization.

3: DT <- get_test_corpora()
- Collect test corpora for validation and comparison; taking a random subset of the full corpus.
- Sources and types similar to domain-specific corpora.
- Ensure diversity and relevance to the domain.

4: C <- build_corpus(SK)
- Apply text processing techniques to refine the corpora (DL).
- Use text processing techniques with the genism library:
  - Remove stop words.
  - Perform stemming and lemmatization.
  - Generate word cloud visualization.
  - Transform text to lowercase.
  - Remove URLs.
  - Tokenize text using Regexp tokenizer.
  - Filter out stop words and numbers.
  - Extract and rank high-frequency words.
- Use word clouds and word groups for further refinement.

5: TM <- LDA(C)
- Converts the preprocessed corpus C into a Document-Term Matrix (DTM) where rows are documents and columns are terms.

Defines hyperparameters like K (number of topics), $\alpha$ (document-topic density), $\beta$ (topic-word density), and number of iterations (*i*).
- Trains LDA Model using the hyperparameters
- Presents resulting topics and their word distributions using matrices.

*Phase 2: Refine – Aspect Weighting from Expert Sources*

6: AText <- Aspect_Identification (External Sources)
- Collect web content from credible sources, including articles, conference materials (keynotes, presentations, proceedings) focused on the domain aspects.
- Clean and preprocess aspect text (remove HTML tags, punctuation, stop words) and define aspects

7: AT <- Weighted_Aspect_Keywords (AText)
- Use TF-IDF to identify and weight top keywords for each aspect. Compute:

$\text{TF-IDF}(t, \text{AText}_i) = \text{tf}(t, \text{AText}_i) \times \log(N / \text{df}(t))$

Select top 50 weighted terms per aspect to serve as semantic anchors.

*Phase 3: Align – Expert-Guided Alignment Layer*

| | |
|---|---|
| // Relevance scores computation - Iterate through each aspect/topic; Cluster the test documents (DT)<br><br>8: ATMi <- ATM (TM, ATi)<br>9: newATMi <- Inferring (DT, ATMi) ∪ newATMi; newTM <- Inferring (TM, DT) | // 8 ATi= Aspect keywords extracted.; Re-weighted topics using aspect keywords ATi by multiplying the terms weights and generate associated Aspect Topic Model ATMi.<br>// 9. For each test document, infer it to initial topic model and each aspect topic model by multiplying the term weights; Aggregate documents inferred in ATMi |

*Phase 4: Stability and Evolution – Iterative Refinement*

// Verify and compare the initial topic model with the aspect topic model on test documents

10: VR <- compare_topic_models(newTM, newATMs, DT)

- Compare newTM with refined aspect models newATMs using test corpora DT

11: fine_tuned_topics = fine_tune_topics(newTM, newATMs, VR)

- Interpret the models to demonstrate how a topic tunned on an aspect with injecting new relevant documents.

**Appendix 7:** Word-Topic Distribution Heatmap: Top Words Across 39 Topics with Aspect 1

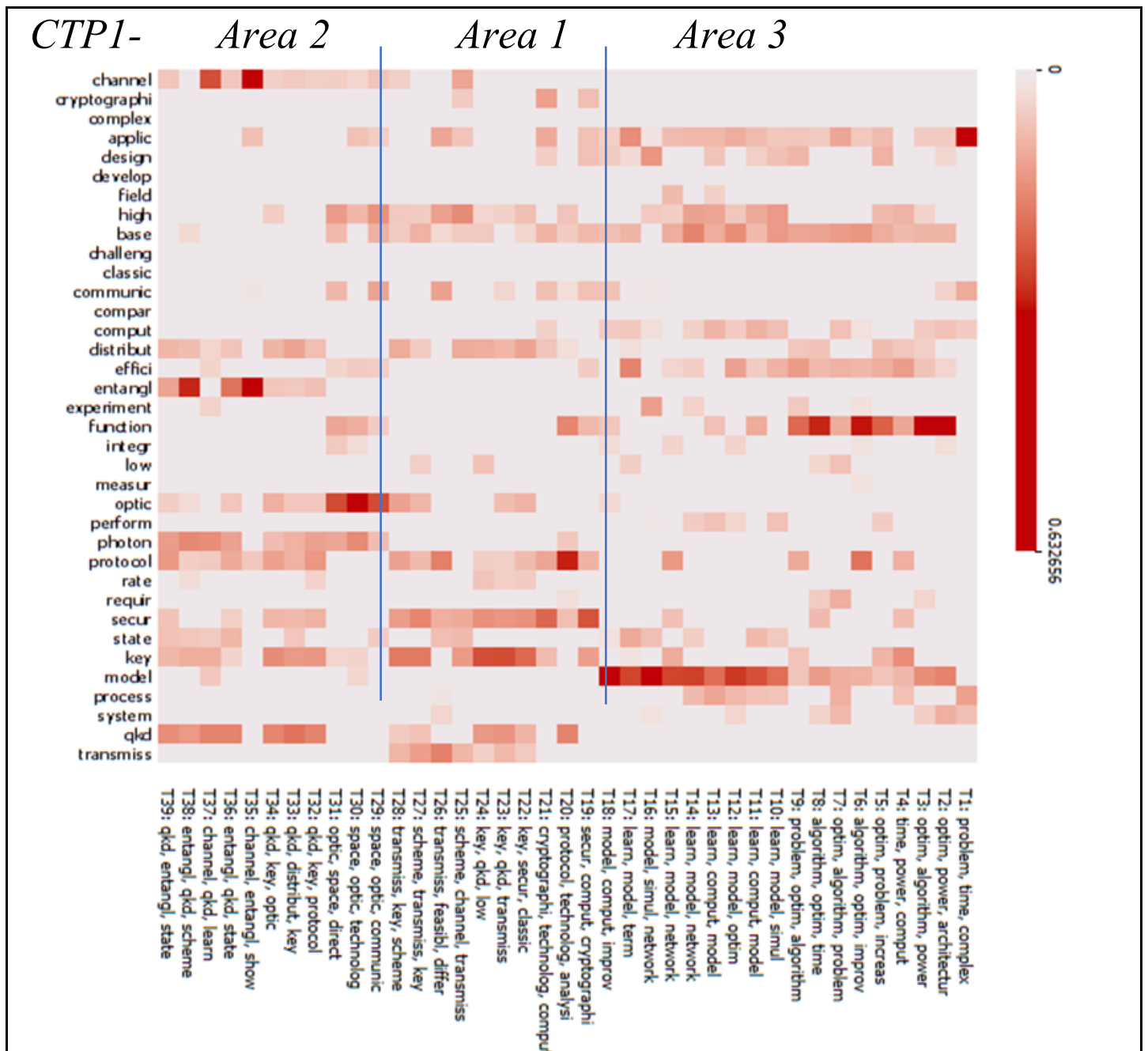

*Aspect 1 – topic model*

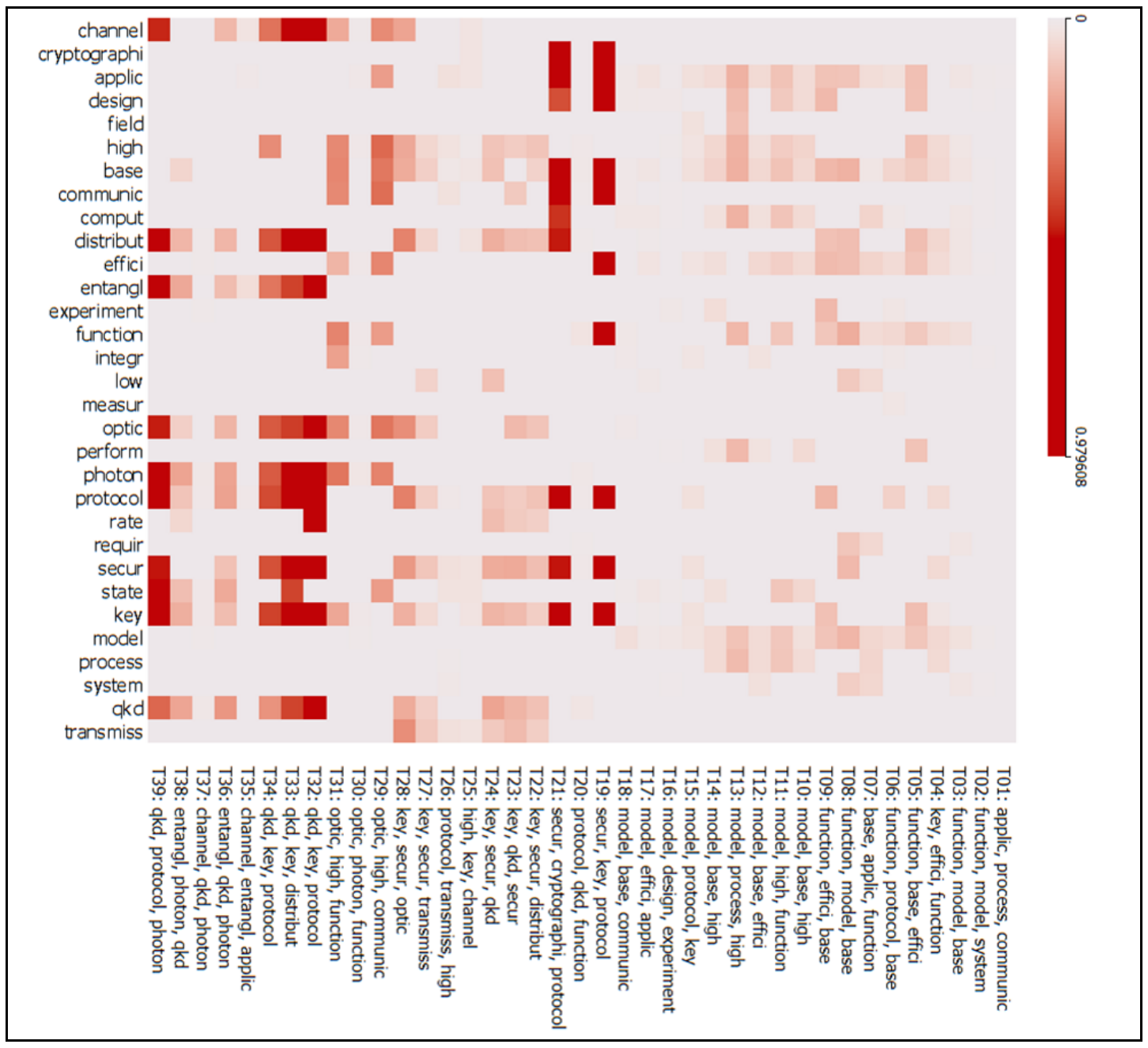

**Appendix 8:** Word-Topic Distribution Heatmap: Top Words Across 39 Topics with Aspect 2

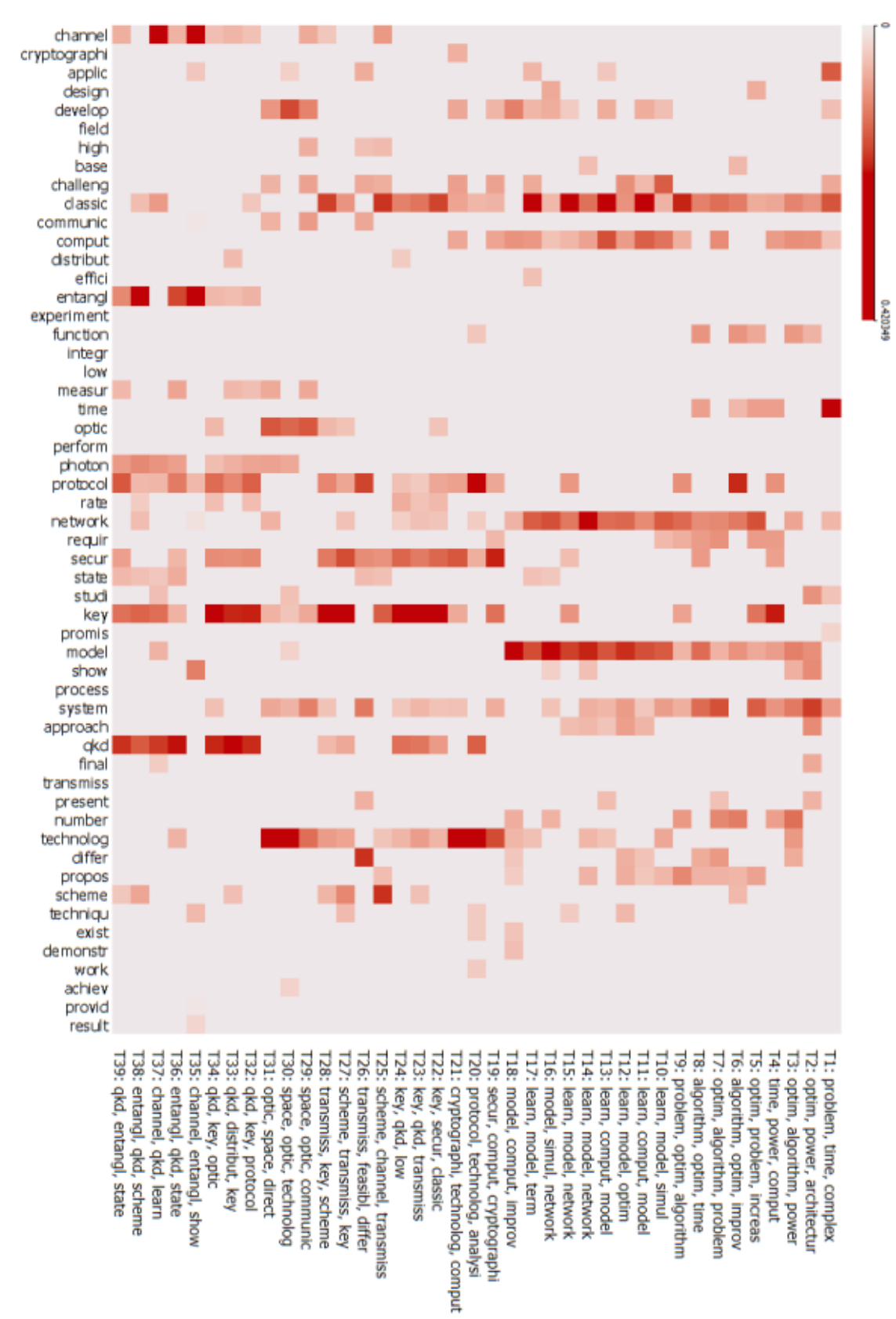

**Appendix 9:** Aspect Identification Process

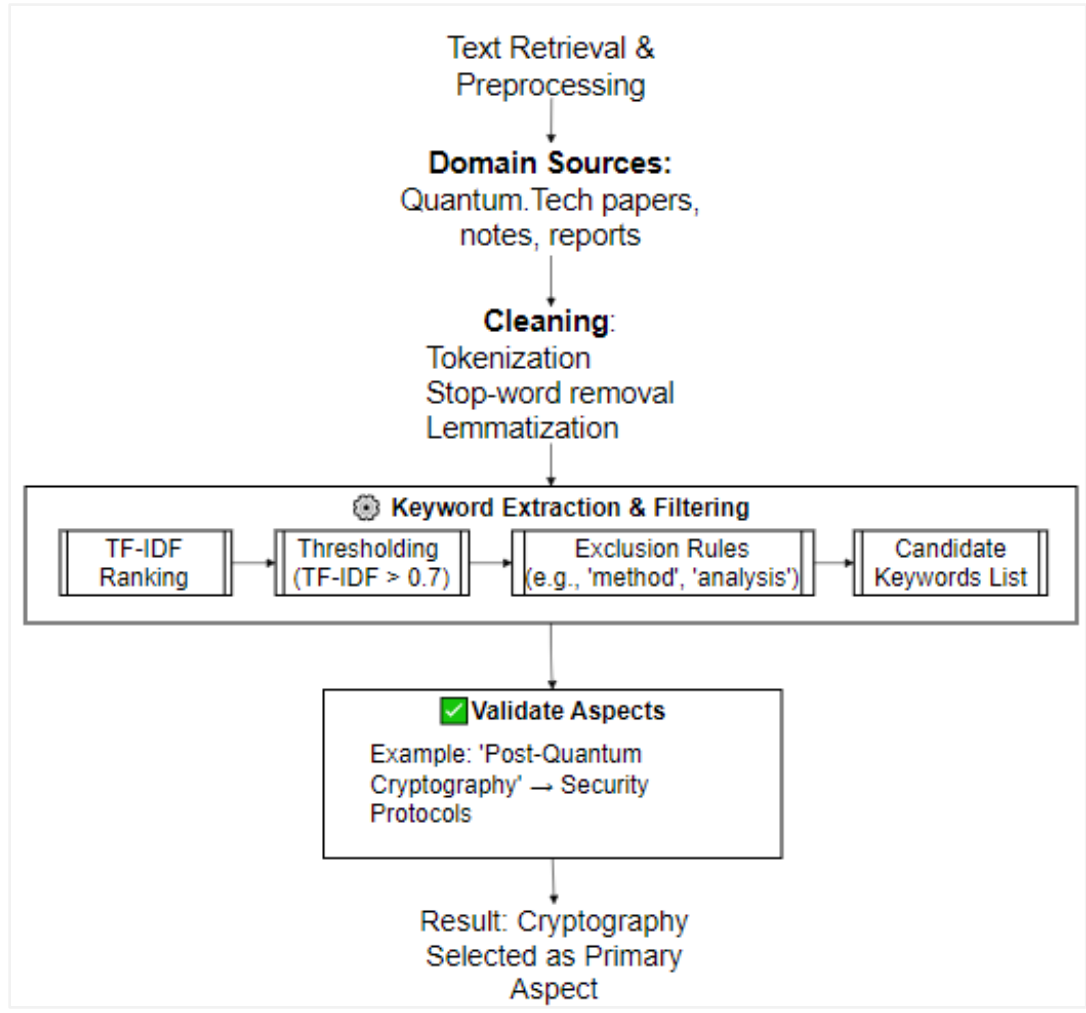

## References

1. Blei DM, Ng AY, Jordan MI. Latent dirichlet allocation. Journal of machine Learning research. 2003;3(Jan):993–1022.
2. Grootendorst M. BERTopic: Neural topic modeling with a class-based TF-IDF procedure. arXiv; 2022, Available from: http://arxiv.org/abs/2203.05794
3. Antons D, Grünwald E, Cichy P, Salge TO. The application of text mining methods in innovation research: current state, evolution patterns, and development priorities. R&D Management. 2020;50(3):329–51.
4. Coccia M, Roshani S. Evolution of topics and trends in emerging research fields: multiple analyses with entity linking, Mann–Kendall test and burst methods in cloud computing. Scientometrics. 2024 Sep;129(9):5347–71.
5. Zhang K, Lian Z, Li J, Li H, Hu X. Short Text Clustering with a Deep Multi-embedded Self-supervised Model. In: Farkaš I, Masulli P, Otte S, Wermter S, editors. Artificial Neural Networks and Machine Learning – ICANN 2021. Cham: Springer International Publishing; 2021. p. 150–61. (Lecture Notes in Computer Science; vol. 12895). Available from: https://link.springer.com/10.1007/978-3-030-86383-8_12
6. Hu Y, Milios EE, Blustein J. Document clustering with dual supervision through feature reweighting. Comput Intell. 2016 Aug;32(3):480–513.
7. Mei JP, Wang Y. Hyperspherical Fuzzy clustering for online document categorization. In: 2016 IEEE International Conference on Fuzzy Systems (FUZZ-IEEE). Vancouver, BC, Canada: IEEE; 2016.
8. Zhao Q, Yang J, Wang Z, Chu Y, Shan W, Tuhin IAK. Clustering massive-categories and complex documents via graph convolutional network. In: Knowledge Science, Engineering and Management. Cham: Springer International Publishing; 2021. p. 27–39. (Lecture notes in computer science).
9. Calof J, Smith J. The integrative domain of foresight and competitive intelligence and its impact on R&D management: Integrative domain of foresight and competitive intelligence. R&D Management. 2009 Dec 16;40(1):31–9.
10. Veugelers M, Bury J, Viaene S. Linking technology intelligence to open innovation. Technological forecasting and social change. 2010;77(2):335–43.
11. Petersen K, Feldt R, Mujtaba S, Mattsson M. Systematic Mapping Studies in Software Engineering. In 2008. Available from: https://scienceopen.com/document?vid=6d552894-2cc3-4e2b-a483-41fa48a37ef8
12. Petersen K, Vakkalanka S, Kuzniarz L. Guidelines for conducting systematic mapping studies in software engineering: An update. Information and Software Technology. 2015 Aug;64:1–18.
13. Blei DM, Lafferty JD. A correlated topic model of science. 2007; Available from: https://projecteuclid.org/journals/annals-of-applied-statistics/volume-1/issue-1/----Custom-HTML----A/10.1214/07-AOAS114.short
14. Sun Y, Li L, Chen Y, Kataev MY. An empirical study on innovation ecosystem, technological trajectory transition, and innovation performance. Journal of Global Information Management (JGIM). 2021;29(4):148–71.
15. Mikolov T, Yih W tau, Zweig G. Linguistic regularities in continuous space word representations. In: Proceedings of the 2013 conference of the north american chapter of the association for computational linguistics: Human language technologies. 2013. p. 746–51. Available from: https://aclanthology.org/N13-1090.pdf
16. Salton G, Buckley C. Term-weighting approaches in automatic text retrieval. Information processing & management. 1988;24(5):513–23.
17. Mei JP, Wang Y, Chen L, Miao C. Incremental fuzzy clustering for document categorization. In: 2014 IEEE International Conference on Fuzzy Systems (FUZZ-IEEE). Beijing, China: IEEE; 2014.

18. Zhuang Y, Gao H, Wu F, Tang S, Zhang Y, Zhang Z. Probabilistic word selection via topic modeling. IEEE Transactions on Knowledge and Data Engineering. 2014;27(6):1643–55.
19. Jagarlamudi J, Daumé III H, Udupa R. Incorporating lexical priors into topic models. In: Proceedings of the 13th Conference of the European Chapter of the Association for Computational Linguistics. 2012. p. 204–13. Available from: https://aclanthology.org/E12-1021.pdf
20. Li C, Chen S, Xing J, Sun A, Ma Z. Seed-guided topic model for document filtering and classification. ACM Trans Inf Syst. 2019 Jan;37(1):1–37.
21. Hu Y, Milios EE, Blustein J. Document Clustering With Dual Supervision Through Feature Reweighting. Computational Intelligence. 2016 Aug;32(3):480–513.
22. Knisely BM, Pavliscsak HH. Research proposal content extraction using natural language processing and semi-supervised clustering: A demonstration and comparative analysis. Scientometrics. 2023 May;128(5):3197–224.
23. Diaz-Valenzuela I, Loia V, Martin-Bautista MJ, Senatore S, Vila MA. Automatic constraints generation for semisupervised clustering: experiences with documents classification. Soft Comput. 2016 Jun;20(6):2329–39.
24. Kong W, Bendersky M, Najork M, Vargo B, Colagrosso M. Learning to cluster documents into workspaces using large scale activity logs. In: Proceedings of the 26th ACM SIGKDD International Conference on Knowledge Discovery & Data Mining. New York, NY, USA: ACM; 2020.
25. Gupta AK, Smith KG, Shalley CE. The Interplay Between Exploration and Exploitation. AMJ. 2006 Aug;49(4):693–706.
26. Seo E, Song ,Jaeyong, and Jin C. Heterogeneity of optimal balance between exploration and exploitation:the moderating roles of firm technological capability and industry alliance network position. Industry and Innovation. 2023 Apr 21;30(4):423–51.
27. Lin Y, Gao X, Chu X, Wang Y, Zhao J, Chen C. Enhancing neural topic model with multi-level supervisions from seed words. In: Findings of the Association for Computational Linguistics: ACL 2023. Stroudsburg, PA, USA: Association for Computational Linguistics; 2023.
28. Jingnan L, Chuan L, Ruizhang H, Yongbin Q, Yanping C. Intention-guided deep semi-supervised document clustering via metric learning. J King Saud Univ - Comput Inf Sci. 2023 Jan;35(1):416–25.
29. Diam TU, Chiavetta D, Porter AL, Saritas O, editors. Anticipating Future Innovation Pathways Through Large Data Analysis. Cham: Springer International Publishing; 2016. (Innovation, Technology, and Knowledge Management). Available from: http://link.springer.com/10.1007/978-3-319-39056-7
30. Hevner AR, March ST, Park J, Ram S. Design Science in Information Systems Research. MIS Quarterly. 2004;28(1):75–105.
31. Peffers K, Tuunanen T, Rothenberger MA, Chatterjee S. A Design Science Research Methodology for Information Systems Research. Journal of Management Information Systems. 2007 Dec;24(3):45–77.
32. Cavaliere F, Prati E, Poti L, Muhammad I, Catuogno T. Secure Quantum Communication Technologies and Systems: From Labs to Markets. Quantum Reports. 2020 Jan 22;2(1):80–106.
33. Manzalini A. Quantum Communications in Future Networks and Services. Quantum Reports. 2020 Mar 11;2(1):221–32.
34. Hassija V, Chamola V, Saxena V, Chanana V, Parashari P, Mumtaz S, et al. Present landscape of quantum computing. IET Quantum Communication. 2020 Dec;1(2):42–8.
35. Settles B. Active Learning Literature Survey. 2009.
36. Manning CD. Introduction to information retrieval. Cambridge university press; 2008. Available from: http://diglib.globalcollege.edu.et:8080/xmlui/bitstream/handle/123456789/1096/Manning_introduction_to_information_retrieval.pdf?sequence=1&isAllowed=y
37. Desai M, Spink A. An algorithm to cluster documents based on relevance. Inf Process Manag. 2005 Sep;41(5):1035–49.
38. Eick CF, Zeidat N, Zhao Z. Supervised clustering-algorithms and benefits. In: 16Th IEEE international conference on tools with artificial intelligence. 2004. p. 774–6.
39. Nguyen DQ, Billingsley R, Du L, Johnson M. Improving topic models with latent feature word representations. Transactions of the Association for Computational Linguistics. 2015;3:299–313.
40. Watanabe K, Baturo A. Seeded Sequential LDA: A Semi-Supervised Algorithm for Topic-Specific Analysis of Sentences. Social Science Computer Review. 2024 Feb;42(1):224–48.
41. Mei JP. Semisupervised fuzzy clustering with partition information of subsets. IEEE Transactions on Fuzzy Systems. 2018;27(9):1726–37.
42. Kuhn KD. Using structural topic modeling to identify latent topics and trends in aviation incident reports. Transp Res Part C Emerg Technol. 2018 Feb;87:105–22.
43. Li X, Li C, Chi J, Ouyang J, Li C. Dataless text classification: A topic modeling approach with document manifold. In: Proceedings of the 27th ACM international conference on information and knowledge management. 2018. p. 973–82.
44. Li N, Luo W, Yang K, Zhuang F, He Q, Shi Z. Self-organizing weighted incremental probabilistic latent semantic analysis. Int J Mach Learn Cybern. 2018 Dec;9(12):1987–98.
45. Liu W, Pang J, Li N, Zhou X, Yue F. Research on Multi-label Text Classification Method Based on tALBERT-CNN. Int J Comput Intell Syst. 2021 Dec;14(1):201.
46. Peinelt N, Nguyen D, Liakata M. tBERT: Topic models and BERT joining forces for semantic similarity detection. In: Proceedings of the 58th annual meeting of the association for computational linguistics. 2020. p. 7047–55. Available from: https://aclanthology.org/2020.acl-main.630/

47. Liu L, Tang L, Dong W, Yao S, Zhou W. An overview of topic modeling and its current applications in bioinformatics. SpringerPlus. 2016 Dec;5(1):1608.
48. Li X, Ouyang J, Zhou X. Supervised topic models for multi-label classification. Neurocomputing. 2015;149:811–9.
49. Wang W, Guo B, Shen Y, Yang H, Chen Y, Suo X. Twin labeled LDA: a supervised topic model for document classification. Appl Intell. 2020 Dec;50(12):4602–15.
50. Magnusson M, Jonsson L, Villani M. DOLDA: a regularized supervised topic model for high-dimensional multi-class regression. Comput Stat. 2020 Mar;35(1):175–201.
51. Li X, Ouyang J, Zhou X. Labelset topic model for multi-label document classification. J Intell Inf Syst. 2016 Feb;46(1):83–97.
52. Li C, Xing J, Sun A, Ma Z. Effective document labeling with very few seed words: A topic model approach. In: Proceedings of the 25th ACM international on conference on information and knowledge management. 2016. p. 85–94.
53. Zha D, Li C. Multi-label dataless text classification with topic modeling. Knowledge and Information Systems. 2019;61:137–60.
54. Li X, Li C, Chi J, Ouyang J, Li C. Dataless Text Classification: A Topic Modeling Approach with Document Manifold. In: Proceedings of the 27th ACM International Conference on Information and Knowledge Management. Torino Italy: ACM; 2018. p. 973–82. Available from: https://dl.acm.org/doi/10.1145/3269206.3271671
55. Wang Y, Jin D, Musial K, Dang J. Community detection in social networks considering topic correlations. In: Proceedings of the AAAI Conference on Artificial Intelligence. 2019. p. 321–8. Available from: https://aaai.org/ojs/index.php/AAAI/article/view/3801
56. Fard MM, Thonet T, Gaussier E. Seed-guided deep document clustering. In: Lecture Notes in Computer Science. Cham: Springer International Publishing; 2020. p. 3–16. (Lecture notes in computer science).
57. Li X, Wang Y, Ouyang J, Wang M. Topic extraction from extremely short texts with variational manifold regularization. Mach Learn. 2021 May;110(5):1029–66.
58. Wang Z, Zheng Y, Zhu H, Yang C, Chen T. Transferable adversarial examples can efficiently fool topic models. Computers & Security. 2022 Jul 1;118:102749.
59. Lin B, Zecevic S, Bouneffouf D, Cecchi G. TherapyView: Visualizing Therapy Sessions with Temporal Topic Modeling and AI-Generated Arts. arXiv; 2023. Available from: http://arxiv.org/abs/2302.10845
60. Zhang Y, Xu FF, Li S, Meng Y, Wang X, Li Q, et al. HiGitClass: Keyword-driven hierarchical classification of GitHub repositories. In: 2019 IEEE International Conference on Data Mining (ICDM). Beijing, China: IEEE; 2019.
61. Ishii M, Matsuura S, Mori K, Tomoishi M, Jin Y, Kitaguchi Y. A study of classification of texts into categories of cybersecurity incident and attack with topic models. In: Proceedings of the 6th International Conference on Information Systems Security and Privacy. Valletta, Malta: SCITEPRESS - Science and Technology Publications; 2020.
62. Mei Q, Ling X, Wondra M, Su H, Zhai C. Topic sentiment mixture: modeling facets and opinions in weblogs. In: Proceedings of the 16th international conference on World Wide Web. Banff Alberta Canada: ACM; 2007. p. 171–80. Available from: https://dl.acm.org/doi/10.1145/1242572.1242596
63. Paul M, Girju R. A two-dimensional topic-Aspect Model for discovering multi-faceted topics. Proc Conf AAAI Artif Intell. 2010 Jul;24(1):545–50.